\documentclass[journal=cmatex,manuscript=article]{achemso}

\usepackage[version=3]{mhchem} 
\usepackage[T1]{fontenc}       

\usepackage{amssymb}
\usepackage{amsmath}

\usepackage{color}

\usepackage[normalem]{ulem}
\usepackage{comment}
\usepackage{soul}

\usepackage{xr}
\makeatletter
\newcommand*{\addFileDependency}[1]{
\typeout{(#1)}
\@addtofilelist{#1}
\IfFileExists{#1}{}{\typeout{No file #1.}}
}\makeatother
\newcommand*{\myexternaldocument}[1]{%
\externaldocument{#1}%
\addFileDependency{#1.tex}%
\addFileDependency{#1.aux}%
}
\myexternaldocument{SI}

\author{Tianyu Su}
\affiliation[A]
{Department of Materials Science and Engineering, University of Illinois at Urbana-Champaign, 1304 W. Green Street, Urbana, Illinois 61801, United States}

\author{Brian J. Blankenau}
\affiliation[B]
{Department of Mechanical Science and Engineering, University of Illinois at Urbana-Champaign, 1206 W. Green Street, Urbana, Illinois 61801, United States}

\author{Namhoon Kim}
\affiliation[B]
{Department of Mechanical Science and Engineering, University of Illinois at Urbana-Champaign, 1206 W. Green Street, Urbana, Illinois 61801, United States}

\author{Jessica A. Krogstad}
\affiliation[A]
{Department of Materials Science and Engineering, University of Illinois at Urbana-Champaign, 1304 W. Green Street, Urbana, Illinois 61801, United States}

\author{Elif Ertekin}
\affiliation[B]
{Department of Mechanical Science and Engineering, University of Illinois at Urbana-Champaign, 1206 W. Green Street, Urbana, Illinois 61801, United States}
\alsoaffiliation[C]
{Materials Research Laboratory, University of Illinois at Urbana-Champaign, 104 South Goodwin Avenue, Urbana, Illinois 61801, United States}
\email{ertekin@illinois.edu}

\title{ First-principles and cluster expansion study of the effect of magnetism on short-range order in Fe-Ni-Cr austenitic stainless steels }

\begin{document}


\newpage
\begin{abstract}
Short-range order (SRO), the regular and predictable arrangement of atoms over short distances, 
alters the mechanical properties of technologically relevant structural materials such as medium/high entropy alloys and austenitic stainless steels.
In this study, we present a generalized spin cluster expansion (CE) model and show that magnetism is a primary factor influencing the level of SRO present in austenitic Fe-Ni-Cr alloys. 
The spin CE consists of a chemical cluster expansion combined with an Ising model for Fe-Ni-Cr austenitic alloys.  
It explicitly accounts for local magnetic exchange interactions, thereby capturing the effects of finite temperature magnetism on SRO. 
Model parameters are obtained by fitting to a first-principles data set comprising both chemically and magnetically diverse FCC configurations. 
The magnitude of the magnetic exchange interactions are found to be comparable to the chemical interactions. 
Compared to a conventional implicit magnetism CE built from only magnetic ground state configurations, the spin CE shows improved performance on several experimental benchmarks over a broad spectrum of compositions, particularly at higher temperatures due to the explicit treatment of magnetic disorder. 
We find that SRO is strongly influenced by alloy Cr content, since Cr atoms prefer to align antiferromagnetically with nearest neighbors but become magnetically frustrated with increasing Cr concentration.  
Using the spin CE, we predict that increasing the Cr concentration in typical austenitic stainless steels promotes the formation of SRO and increases order-disorder transition temperatures.
This study underscores the significance of considering magnetic interactions explicitly when exploring the thermodynamic properties of complex transition metal alloys. 
It also highlights guidelines for customizing SRO  through adjustments of alloy composition. 
\end{abstract}

\newpage 

\section{Introduction}

Short-range order (SRO), the ordered arrangement of atoms over limited distances~\cite{cohen1962some}, affects the mechanical behavior~\cite{zhang2020short,yin2020yield,ding2018tunable}, magnetic transitions~\cite{kollie1973heat, cadeville1987magnetism}, electronic transport~\cite{tulip2008theory,mu2018electronic}, and lattice dynamics~\cite{rucker1996phonons,alam2011phonon} of functional structural alloys. 
The existence of SRO in alloys was first determined by diffuse X-ray scattering~\cite{cohen1962some,cowley_1965_shortrange}, and later verified by neutron diffraction~\cite{cenedese1984diffuse,menshikov1997local,zhang2017local}, transmission electron microscopy~\cite{ding2019tuning,zhang2019direct,zhang2020short,chen2021direct}, and atom probe tomography~\cite{li2023quantitative}.
While the experimental characterization of SRO poses ongoing challenges, advances in state-of-the-art characterization techniques are progressively enhancing the understanding of its chemical and structural properties.

The impact of SRO on the mechanical properties of technologically important structural materials like medium and high entropy alloys (MEA, HEA) and austenitic stainless steels are becoming better understood both experimentally\cite{zhang2020short,yin2020yield,kim2022effects,hong2023investigation} and theoretically\cite{ding2018tunable,li2019strengthening,pei2020statistics,yin2021atomistic,kang2014impact,chu2023investigation}.
Nevertheless, there exists a diversity of perspectives on how and to what extent SRO affects the mechanical performance of structural materials. 
Experiments reported by \citeauthor{zhang2020short} indicated that SRO enhances the hardness and yield strength of CrCoNi MEA~\cite{zhang2020short}.
However, others have argued that SRO may be negligible under typical processing conditions, thus reducing its influence on misfit volume and alloy hardness~\cite{yin2020yield}. \citeauthor{ding2018tunable} theoretically quantified the impact of SRO on stacking fault energy (SFE) and identified a correlation between increasing degrees of SRO and higher SFE values~\cite{ding2018tunable}.
In other computational studies, it was reported that SRO can raise activation barriers for dislocation motion, thus affecting dislocation mobility~\cite{yin2021atomistic,li2019strengthening}.
For austenitic stainless steels, strengthening effects due to SRO have been suggested from molecular dynamics (MD) simulations~\cite{chu2023investigation}. 
In face-centered cubic (FCC) alloys, SRO is believed to promote planar slip and dislocation pile-ups due to glide plane softening~\cite{gerold1989origin}.
These varied viewpoints showcase the multifaceted nature of SRO, and the need for further exploration to reconcile differing perspectives. 

From an atomistic point of view, SRO is promoted by attractive or repulsive chemical interactions between elemental species. 
Recent studies have shown that magnetic interactions can be a driving force for atomic-scale ordering in NiFeCrCo HEA and CrCoNi MEA~\cite{niu2015spin,walsh2021magnetically}. 
It has also been shown that temperature-induced magnetization reduction has a strong impact on effective chemical interactions in Fe-Ni alloys~\cite{ekholm2010influence}.
For Fe-Ni-Cr alloys, the influence of longitudinal spin fluctuations is nontrivial, as disordered magnetism at high temperatures has been shown to reduce ordering tendencies~\cite{ruban2016atomic}.
Similar trends were discovered in other transition metal alloys such as CrCoNi and CrMnFeCoNi, where magnetic states exert substantial effects on SRO~\cite{woodgate2023interplay}. 
On the other hand, ~\citeauthor{ghosh2022short} have argued that magnetism is not responsible for the ground-state chemical ordering in Cr-Co-Ni alloys~\cite{ghosh2022short}, albeit via MD simulations that did not include finite-temperature magnetism.
The importance of magneto-structural coupling has been emphasized in other MD simulations showing that the influence of magnetic degrees of freedom can be non-negligible~\cite{chu2023investigation}.

Due to the inherently small length scale of SRO, experimental examination is challenging and often unfeasible. 
Consequently, atomic-scale simulations are useful for exploring the presence, causes, and implications of SRO.
Atomistic Monte Carlo (MC) simulation~\cite{metropolis_1949}, intended to generate representative thermodynamic configurations by sampling the Boltzmann distribution, is an efficient tool for statistically probing the presence or absence of chemical SRO.
Nonetheless, tens of thousands of energy evaluations are needed for robust statistics to estimate expectation values.
This can be computationally expensive when employing first-principles methods such as density functional theory (DFT). 
To address this computational challenge, effective models based on the cluster expansion (CE) formalism are frequently utilized~\cite{Sanchez1984}.
The CE method provides an approximate yet efficient method to obtain the configurational energy of multi-component systems based on lattice models~\cite{wolverton1994cluster,van2009multicomponent}, and is often combined with MC simulations to study thermodynamic properties~\cite{van2002alloy,van2002automating}.
Warren-Cowley parameters~\cite{cowley_1965_shortrange} and other measures of SRO can be predicted for ternary systems and beyond, such as Fe-Ni-Cr and Fe-Ni-Cr-Mn alloys, using CE-MC simulations~\cite{wrobel2015phase,fedorov2020phase}.
Most typically, CE formulations include chemical contributions but either ignore magnetic degrees of freedom altogether or only account for magnetism implicitly, potentially leading to an overestimation of SRO~\cite{ruban2016atomic}.
Other approaches incorporating magnetism~\cite{ekholm2010influence,ruban2016atomic,woodgate2023interplay} introduce temperature- and composition-dependent pair interactions into the CE framework, making it challenging to extend to diverse alloy systems.

In this work, we demonstrate a generalized spin CE and use it to study the interplay between magnetism and SRO in Fe-Ni-Cr austenitic stainless steels, structural alloys with excellent corrosion resistance and formability. 
The spin CE combines a cluster expansion, where clusters are defined explicitly by the chemical species, together with a spin Ising model to capture magnetic effects.
The spin CE is fitted to first-principles energies of alloys of varying chemical and magnetic states. 
The magnetic exchange interactions exhibit comparable magnitude to the chemical interactions, underscoring the significance of magnetism.  
Both the spin CE and a conventional implicit magnetism CE are compared to available experimental results, including measured Warren-Cowley SRO parameters, order-disorder transition temperatures, and Curie temperatures, showcasing the benefits of accounting for magnetism. 
Using the spin CE, we predict the degree of SRO present in prototype austenitic steels at different temperatures and demonstrate a significant, magnetically-mediated influence of alloy composition on SRO.
By explicitly considering magnetic interactions, this work provides an atomic-scale understanding of SRO formation in austenitic stainless steels. 
Moreover, it highlights potential routes to customize SRO by adjusting alloy composition. 

\section{Methods}

\subsection{Cluster expansion}

Traditionally, cluster expansion (CE) models are designed to parametrize any function of configuration using a set of orthogonal basis functions. 
This parametrization is often considered as a generalized Ising model~\cite{Sanchez1984}.
In a CE, the real alloy system is represented by a lattice model with different species occupying each site.
A spin-like occupation variable $\sigma_i$ is assigned to the occupied site and a particular arrangement of occupations is called a configuration.
Quantities of interest, such as total energies, are then parametrized as a function of site occupation variables, i.e. 
\begin{equation}
    E(\vec{\sigma}) = \sum_{\omega} m_{\omega} J_{\omega} \langle \Gamma_{\omega'} (\vec{\sigma}) \rangle_{\omega} \hspace{1em}. \label{ce1}
\end{equation} 
Here, $\vec{\sigma}$ indicates a given configuration and $E(\vec{\sigma})$ is the energy of configuration $\vec{\sigma}$ per atom/lattice site.
The sum is over all clusters $\omega$ (a set of sites in the lattice) that are symmetry-distinct.
The symbol $m_{\omega}$ is the multiplicity, indicating the number of symmetry-equivalent clusters of type $\omega$.
The fitted parameter $J_{\omega}$ denotes the effective cluster interaction (ECI) for cluster $\omega$, which contains information regarding the energetics of the target system. 
Cluster functions $\Gamma_{\omega} (\vec{\sigma})$ are typically defined as the product of orthonormal point functions of occupation variables $\sigma_i$ over the sites within cluster $\omega$. 
The choice of cluster functions can be found in Supplementary Information(SI).
The average $\langle \Gamma_{\omega'} (\vec{\sigma}) \rangle_{\omega}$ is taken over all clusters $\omega'$ that are equivalent by symmetry to cluster $\omega$.
These cluster functions, together with the ECIs, formally represent the point, pair, and many-body interactions based on the generalized Ising model.
Although the effective Hamiltonian in Equation (\ref{ce1}) is only complete when all possible clusters $\omega$ are considered in the sum, in practice the sum often converges quickly. 
Typically only a finite number of clusters are needed to map CE energies to DFT energies~\cite{van2002automating}.

When the system contains more than two elements, a transformation depending on the cluster functions is needed to obtain pair interactions between atomic species~\cite{wolverton1994cluster,wrobel2015phase} to physically interpret the ECIs in the conventional CE.
When it comes to triplet and higher-order interactions, the challenge of extracting meaningful insights from the ECIs is exacerbated, as disparate interactions are often amalgamated into a single ECI value~\cite{kim2022multisublattice}. 
Using the approach outlined in Ref.~\cite {kim2022multisublattice}, here we introduce a modified CE designed to disentangle meaningful interactions that are challenging to extract from conventional CE. 
The adaptation builds upon the foundations of the conventional approach while eliminating the necessity to use an orthonormal basis set. 
Rather than being decorated with nonlinear point functions, the clusters are decorated with atomic species directly. 
Additionally, we incorporate a spin-1 Ising model into the modified CE to capture the effects of magnetism in the Fe-Ni-Cr configuration space.
Within this model, spin variables take values of +1, 0, or -1 for all elements.
We set element-specific spin thresholds and convert the real magnetic moments (in Bohr magnetons) from DFT to spin variables in CE (see below). 

The Hamiltonian used here is given by a sum of chemical and magnetic interactions:
\begin{equation}
    E_{CE}(\vec{\sigma}) = \sum_{\alpha} J_{\alpha} \Theta_{\alpha} (\vec{\sigma}) + \sum_{\beta} \sum_{\langle i,j \rangle} J_{\beta} S_{i} S_{j} \hspace{1em}. \label{ce2}
\end{equation}
The first term on the right-hand side of Equation (\ref{ce2}) describes the chemical interactions. 
A cluster $\alpha = \{ \alpha_1,...,\alpha_l \}$ of length $l$ is defined as a motif decorated with specific chemical species (no longer a product of orthonormal point functions).
The effective cluster interaction of cluster $\alpha$ is given by $J_{\alpha}$,  and $\Theta_{\alpha}(\vec{\sigma})$ is the occurrence of cluster $\alpha$ appearing in the given configuration $\vec{\sigma}$.
The second term describes the magnetic exchange interactions by summing over all pair
sites $\langle i,j \rangle$ within a given distance.   
The magnetic exchange interaction of magnetic dimer $\beta$ is given by $J_{\beta}$.
In full generality, Equation (\ref{ce2}) contains a large number of chemical and spin clusters limited only by the size of the selected cutoff radius. 
Down selection of the clusters is achieved by using a compressive sensing method~\cite{nelson2013compressive} where the optimal clusters and their ECIs are obtained from
\begin{equation}
    J = arg \min_{J} \left\{ \frac{1}{N} \sum_{i=1}^N ( E_{i, DFT} - E_{i, CE}(\vec{\sigma}) )^2 + \lambda \sum \left| J \right| \right\} \hspace{1em}.
	\label{lasso}
\end{equation}
Leveraging the Least Absolute Shrinkage and Selection Operator (LASSO)~\cite{tibshirani1996regression} in Equation (\ref{lasso}), we obtain clusters that offer the highest predictive power. 
To avoid both overfitting and underfitting, we use a 10-fold cross-validation (CV) for all fitting procedures and incorporate regularization parameter $\lambda$ when evaluating the corresponding set of ECIs. 
This model, referred to here as the spin CE, explicitly accounts for magnetic degrees of freedom in the alloy, treating both chemical and magnetic interactions on equal footing.

In contrast to Equation (\ref{ce2}),  many CE models for magnetic alloys simplify the effects of spin by (i) neglecting the second term of Equation (\ref{ce2}), and (ii) including only the lowest energy spin configurations for a given chemical ordering when fitting ECI~\cite{wrobel2015phase,fedorov2020phase,fedorov2023composition}.
That is, the CE model is built with chemistry terms only and, given a chemical configuration, it is assumed that the spin degrees of freedom relax to their magnetic ground state (at all temperatures). 
In this approach, the effects of magnetism are incorporated only implicitly rather than explicitly, and it is assumed that chemical interactions dominate and finite temperature magnetism introduces negligible effects. 
We also consider this approach, referred to here as the implicit magnetism CE, and compare it to the spin CE.   

\subsection{Monte Carlo}

A lattice Monte Carlo (MC) method in the canonical ensemble was implemented based on the modified CE formalism.
Kawasaki dynamics~\cite{kawasaki1966diffusion} for atom swaps was used to ensure that the composition of the system remained fixed.
Atom swaps and spin-flips were enabled for each MC step with equal probabilities to allow simultaneous configurational and magnetic equilibration. 
We considered temperatures between 600 K and 1500 K in 100 K increments. 
To evaluate quantities of interest, the structures were initialized as disordered configurations, equilibrated at the highest temperature, and then cooled down and equilibrated at each subsequent temperature. 
The Metropolis-Hastings algorithm~\cite{metropolis_1953} was used to evaluate expectation values sampled from the Boltzmann distribution.
We used a 10$\times$10$\times$10 conventional FCC supercell containing 4000 atoms, for which energies are converged to within 0.1 meV/atom, to exclude finite-size effects. 
For each temperature of interest, we set 2000 MC steps per atom for equilibration, followed by 8000 passes for the evaluation of thermodynamic quantities, for each temperature.
Convergence tests were performed to ascertain that the number of passes was sufficient for the system to reach equilibrium also to within 0.1 meV/atom. 

For model evaluation, the Curie temperature and chemical order-disorder transition temperature were obtained from the temperature dependence of the specific heat.
The Warren-Cowley SRO parameter 
\begin{equation}
    \alpha_l^{AB} = 1 - \frac{P_l^{AB}}{C_AC_B} = 1 - \frac{p_{l,A}^B}{C_B} \hspace{0.5em}, 
\label{WC_SRO}
\end{equation} 
was determined by sampling the pair probability.
Here $P_l^{AB}$ is the probability of finding $AB$ pairs in the $l$-th neighbor shell, and $p_{l,A}^B$ = $P_l^{AB}$/$C_A$ is the conditional probability of finding atom $B$ in the $l$-th coordination shell of atom $A$. 
The symbols $C_A$ and $C_B$ are the concentration of $A$ and $B$ atoms, respectively. 
If there is no correlation between $A$ and $B$, as in a random solution, then $\alpha$ vanishes since $P_l^{AB}$ = $C_AC_B$. 
A preference for like-pairs ($AA$ and $BB$ clustering) is given by $\alpha > 0$, while $\alpha < 0$ indicates an ordering tendency for unlike pairs ($AB$ ordering).

\subsection{First-principles data generation}

Spin-polarized density functional theory (DFT) calculations were performed using the Projected Augmented Wave (PAW) method~\cite{kresse1999ultrasoft}, as implemented in the Vienna \textit{Ab-initio} Simulation Package (VASP)~\cite{kresse1993ab,kresse1996efficient}.  
We used the Perdew-Burke-Erzenhof (PBE) approximation to the exchange-correlation functional~\cite{perdew1996generalized}.  
PAW-PBE pseudopotentials were employed with semi-core states frozen. 
The atomic configurations for Fe, Ni, and Cr were [Ar]3d$^7$4s$^1$, [Ar]3d$^9$4s$^1$, and [Ar]3d$^5$4s$^1$, respectively. 
In all cases, the plane wave cutoff was set to 500 eV. 
For $k$-point sampling, we use a density of 2400 $k$-points per reciprocal atom. 
This corresponds to a $11 \times 11 \times 11$ Monkhorst-Pack $\Gamma$-centered mesh for a single atom FCC unit cell, and scales inversely with increasing cell size. 
Testing was performed to ensure $k$-point sampling convergence within 0.4 meV/atom.
Fermi-level smearing was applied using the first-order Methfessel-Paxton method, with a smearing width of 0.05 eV.
During geometry optimization, the energy precision was set to $10^{-6}$ eV/cell, with all the forces between atoms converged to be less than 0.02 eV/\AA. 

In this study, we focus on the FCC Fe-Ni-Cr alloys relevant to austenitic stainless steels.
The DFT calculations were employed to generate a two-part data set from which we could construct a CE model.
The first part contains 237 unit cell structures generated automatically by the Alloy Theoretic Automated Toolkit (ATAT)~\cite{van2002alloy} via a variance reduction scheme for the ternary Fe-Ni-Cr system. 
The second part encompasses 202 2$\times$2$\times$2 FCC special quasi-random structures (SQS) of varying Fe-Ni-Cr compositions, generated using the \texttt{mcsqs} code~\cite{van2013efficient}.  
These structures were systematically generated to sample the ternary composition space uniformly. 
Detailed alloy compositions are provided in SI Figure~\ref{composition_space}.

The importance of accurately accounting for magnetism in alloys such as Fe-Ni-Cr, especially to obtain representative DFT predictions of energies in the composition space, is well-documented~\cite{wrobel2015phase,ruban2016atomic}. 
To generate a data set that is diverse in terms of magnetism as well as chemical ordering, for each chemical configuration we initialized multiple distinct magnetic spin configurations and carried out DFT relaxations. 
Following an approach similar to Ref.~\cite{wrobel2015phase}, several magnetic states were initialized for each atomic configuration, with the magnetic moments of Fe, Ni, and Cr selected from $\pm5$, $\pm3$, and $\pm3$ $\mu_B$, respectively. 
Since Cr atoms tend to align antiferromagnetically (AFM) with neighboring atoms, particularly Fe atoms~\cite{klaver2006magnetism}, we included initializations in which Fe and Ni atoms were ferromagnetically (FM) aligned, and Cr atoms were given opposite spins. 
This particular initialization often resulted in lower energy configurations when compared to other magnetic arrangements. 
To expand the data set to encompass more diverse magnetic configurations, we also initialized calculations for supercell structures in both fully FM and AFM states.  
We note that the final magnetic configurations themselves change during the simulation, relaxing into nearby local minima of the potential energy surface, which makes it challenging to uniformly sample diverse magnetic configurations. 
For instance, even when initially set as ferromagnetic, Cr atoms most often eventually acquire negative magnetic moments in random Fe-based alloys~\cite{korzhavyi2009electronic}. 
With the inclusion of magnetic degrees of freedom, the total number of unique structures in the data set expanded from 439 to 533 (see SI Figures 2, 3, and 4 for discussion of magnetic configuration diversity). 
When fitting the spin ECI in Equation (\ref{ce2}), the spin thresholds for Fe, Ni, and Cr atoms were determined by inspection of the magnetic moments distributions from DFT, as shown in SI Figure~\ref{mag_distrib}.
Ultimately, the threshold for Fe was set to 1.5 $\mu_B$, for Ni 0.4 $\mu_B$, and for Cr 0.8 $\mu_B$.

\section{Results}

\subsection{The magnetism of Fe-Ni-Cr alloys}

\begin{figure*}[!hbtp] 
\centering
\includegraphics[width=3in]{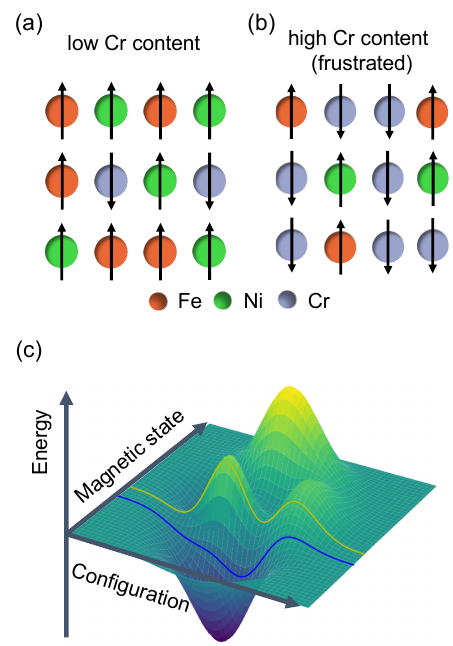}
\caption{
Two-dimensional schematic illustration of the spin configurations when (a) Cr concentration is low and isolated Cr atoms can align antiferromagnetically with neighbors to achieve a favorable spin configuration, and (b) Cr concentration is high, resulting in magnetic frustration. c) Schematic showing energy as a function of magnetic and configurational degrees of freedom. Blue and yellow lines show how energy might change with configuration at different magnetic states.}
\label{fig1} 
\end{figure*}

We begin by investigating the magnetic states of FCC Fe-Ni-Cr alloys. 
Consistent with prior studies~\cite{klaver2006magnetism, korzhavyi2009electronic,wrobel2015phase}, we found that Fe and Ni atoms most typically exhibit a preference for FM alignment with neighbors, while Cr atoms prefer AFM alignment. 
For Cr species, SI Figure~\ref{CrFe} shows the distribution of Cr magnetic moments and the spin products of all Fe-Cr first nearest neighbors (1 NN) present in the full data set.
The distributions are shown separately for Cr atoms with fewer ($\le5$) or more ($\ge6$) nearest neighbor Fe species. 
When a larger number of Fe neighbors are present, the Cr atoms exhibit more negative magnetic moments and the Fe-Cr spin products are statistically more negative. 
This change in magnetization arises because the favorable arrangement of AFM Cr is only possible when the Cr concentration is low as illustrated schematically for a 2D square lattice in Figure~\ref{fig1}(a). 
As the Cr concentration increases, Cr atoms necessarily become nearest neighbors with each other.  
In this case, it is geometrically not possible to arrange Cr atoms to be fully AFM to all neighbors as shown in Figure~\ref{fig1}(b), leading to magnetic frustration. 
Further analysis in SI Figure~\ref{CrCr}(a-c) demonstrates a transition from the AFM (magnetic moment $\le -0.8 \mu_B $) to the nonmagnetic (NM, magnetic moment $\mu_B > -0.8$) state of Cr when the number of neighboring Cr atoms surpasses six, half of the 1 NN coordination number. 
This transition arises from the challenge of achieving global magnetic order among Cr atoms when local Cr concentration is high.  
Intuitively, we expect that this tendency may induce Cr atoms to be spatially separated from each other, in order to achieve energetically favorable spin configurations.
In contrast, this trend is not observed for Fe as shown in SI Figure~\ref{CrCr}(b), which we attribute to a strong preference for FM interactions between Fe and Ni. 

To investigate how these exchange interactions affect alloy stability, SI Figure~\ref{mag_vs_conf} shows the results of non spin-polarized DFT calculations on several alloy compositions such as Fe$_{3}$Ni, FeNi, and others. 
For each composition, we considered various chemical configurations, including $L1_0$, $L1_2$, or random structures.
The corresponding energies are marked with large circles in SI Figure~\ref{mag_vs_conf}.  
The spread in the energies lies between 0.05-0.1 eV/atom, illustrating how much total energies vary with chemical ordering. 
Subsequently, we selected the chemical configurations with the lowest and highest energies, and initialized them with different magnetic states for spin-polarized DFT calculations. 
These energies are also shown in SI Figure~\ref{mag_vs_conf}, labeled by small triangles and diamonds.
For certain compositions (e.g. Cr$_{3}$Ni and Ni$_{2}$Cr), the introduction of magnetism only slightly affects total energies, and preserves the ordering of lowest and highest energy chemical configurations. 
However, for compositions like Ni$_3$Fe and FeNi, magnetic interactions reduce total energies substantially and play a key role in stabilizing the structures.
Additionally, when spin is included for these two compositions, the relative ordering of the highest and lowest chemical configurations becomes reversed. 
These observations highlight the way that magnetism itself may dramatically affect chemical short-range order, and the importance of explicitly including spin degrees of freedom in effective models.  

For example, Figure~\ref{fig1}(c) schematically depicts alloy energy as a function of chemical configuration and magnetic state. 
The energy for a given alloy configuration can change significantly due to the influence of magnetism, and certain configurations may only be stabilized for a specific magnetic state.
Intuitively, magnetically favorable interactions can enhance ordering tendencies distinct from SRO that would be present in the absence of magnetism. 
For example, frustrated magnetic interactions between adjacent Cr atoms may exert a destabilizing impact, prompting Cr atoms to spatially segregate from one another. 
On the other hand, at high enough temperatures, frustrated magnetism may be insufficient to promote certain configurations, leading to alterations in the ordering tendencies.

\subsection{Construction of the CE models}



Having analyzed trends in exchange interactions and their effect on alloy energies, it is now possible to construct a comprehensive CE model that incorporates both configurational and magnetic degrees of freedom. 
The clusters that appear in Equation (\ref{ce2}) include a large set of chemical dimers, trimers, and quadrumers, as well as magnetic dimers. 
To select the clusters that most correlate to configuration energies, we performed several LASSO-CV tests with a systematically expanding set of clusters.
This strategy helps avoid both under and over -fitting, by comparing the CV score or root mean square error (RMSE) for different cluster choices~\cite{van2002alloy,zarkevich2004reliable,kim2022multisublattice}. 
SI Figure~\ref{changing_clusters} illustrates the effect of changing the number and type of clusters on the RMSE, and shows that the minimum RMSE appears in a shallow basin around 7-8 dimers, 8-12 trimers, and 1-3 quadrumers. 

Given the shallow basin in SI Figure 6 where the minimum RMSE is found, we  
selected several benchmark tests and systematically assessed candidate models within the basin indicated by the black circle in SI Figure~\ref{changing_clusters}(b). 
The experimental benchmarks used for evaluation and final model selection are: 
\begin{itemize}
\item SRO parameters for various ternary Fe-Ni-Cr alloys~\cite{menshikov1997local,cenedese1984diffuse}, Fe-Ni binary alloys~\cite{goman1971fine,menshikov1972FeNi} summarized in Ref~\cite{rossiter1984phase}, and Ni-Cr binary alloys~\cite{caudron1992situ,schonfeld1988short}; 
\item order-disorder transition temperatures of FeNi and Ni$_3$Fe alloys~\cite{massalski1990binary,mandal2022l10}; and 
\item Curie temperatures of Ni, FeNi $L1_0$ alloy, and Ni$_3$Fe $L1_2$ alloy~\cite{chicinas2002magnetic,wasilewski1988magnetic}.
\end{itemize}
The best model that matches almost all benchmarks comprises 7 dimers, 12 trimers, and 1 quadrumer, and includes magnetic interactions up to the 3rd nearest neighbors (3 NN) before LASSO selection.
The geometries of all possible chemical clusters are schematically depicted in SI Figure~\ref{all_clusters}.

\begin{figure*}[!hbtp] 
\centering
\includegraphics[width=3in]{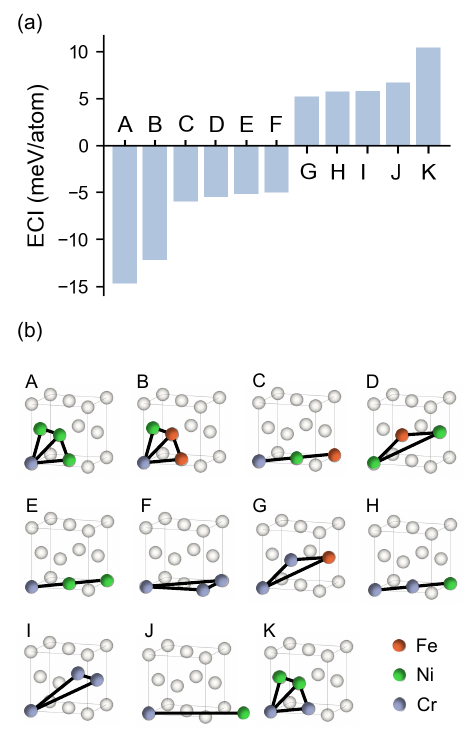}
\caption{ 
(a) The important chemistry ECIs with magnitudes larger than 5 meV/atom. The atomic configurations of the selected chemistry clusters are depicted in (b). Fe, Ni, and Cr atoms are marked with red, green, and blue, respectively
}
\label{fig2}
\end{figure*} 


To visualize the clusters selected in the final model, Figure~\ref{fig2} shows all chemical ECIs with magnitudes exceeding 5 meV/atom. 
Small dimers are found to be excluded from the features selected by LASSO CV, as they are absorbed into trimers that contain nearest-neighbor interactions. 
A consistent pattern is apparent from Figure~\ref{fig2}. 
Positive (unfavorable) ECIs predominantly involve Cr-Cr 1 NN interactions, while the negative (favorable) ECIs exhibit Ni-(Fe/Ni/Cr) 1 NN interactions or Cr-Cr 2nd nearest-neighbor (2 NN) interactions.
In other words, Cr species prefer to bond with Fe or Ni as 1 NN, in agreement with the negative 1 NN Fe-Cr and Ni-Cr SRO parameters obtained previously in Ref.~\cite{cenedese1984diffuse}. 
Ni-containing clusters yield lower energies, consistent with findings that Ni acts as an austenite stabilizer in stainless steels~\cite{ishida1974ferrite}.

\begin{figure*}[!hbtp] 
\centering
\includegraphics[width=3in]{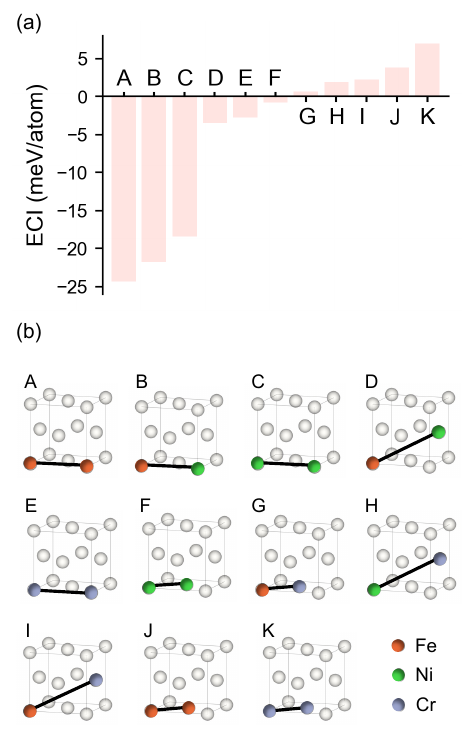}
\caption{ 
(a) The important spin ECIs with magnitudes larger than 0.5 meV/atom. The atomic configurations of the selected spin clusters are depicted in (b). Fe, Ni, and Cr atoms are marked with red, green, and blue, respectively
}
\label{fig3}
\end{figure*}

Figure~\ref{fig3} gives an overview of all magnetic interactions with magnitudes exceeding 0.5 meV/atom. 
Of particular note, some spin interactions exhibit similar or even larger magnitude ECI than chemical interactions, emphasizing the strength of magnetic interactions in Fe-Ni-Cr.
Among the magnetic interactions, the most negative (favorable) ECI are Fe-Fe, Fe-Ni, and Ni-Ni 2 NNs. 
In contrast, Fe-Fe and Cr-Cr 1 NN interactions exhibit the largest positive ECI, signaling their AFM preference. 
In between, there are several ECIs involving all chemical species that show comparatively small magnitude. 
Overall, the ECIs for magnetic interactions obtained from the spin CE are aligned with prior theoretical investigations~\cite{ruban2016atomic,wrobel2015phase,sun2020magnetic}.  

The AFM nature of Fe-Fe 1 NN pairs in the FCC structure has also been reported in Ref.~\cite{ruban2005origin}. 
In that work, depending on the local environment, the 1 NN Fe-Fe magnetic exchange interaction obtained from Green's function method and magnetic force theorem~\cite{szilva2023quantitative} was found to vary between negative and positive values. 
This finding is also consistent with prior findings that the magnetic ground state of FCC Fe in DFT exhibits an AFM double-layer configuration~\cite{wrobel2015phase,sun2020magnetic}. 
However, the magnitude of the Fe-Fe 1 NN interaction is small compared to the magnitude of the FM Fe-Fe 2 NN and FM Fe-Ni 2 NN interaction, which means that the latter interactions will dominate the observed behavior.  
The smaller magnitude of Fe-Fe 1 NN magnetic interactions compared to Fe-Fe 2 NN interactions was also found in Ref.~\cite{ruban2005origin}. 

By contrast, the Cr-Cr 1 NN interaction has the largest magnitude among all interactions involving Cr species.
The magnetic interactions between Cr and Ni (and Cr and Fe) are relatively small, compared to other interactions.
This agrees with the DFT results in SI Figure~\ref{mag_vs_conf}, showing  
that Ni-Cr alloys have similar energies in spin-unpolarized and spin-polarized DFT calculations.
The dominant 1 NN Cr-Cr interaction for Cr species is the reason for the behavior shown in Figure 1(b) and SI Figure 4(a-c), in which the magnetic moments on Cr species transition from AFM to near zero as the number of Cr neighbors increases and the frustration effect becomes significant. 

\subsection{Evaluation of spin CE's predictive performance}

\begin{figure*}[!hbtp] 
\centering
\includegraphics[width=6in]{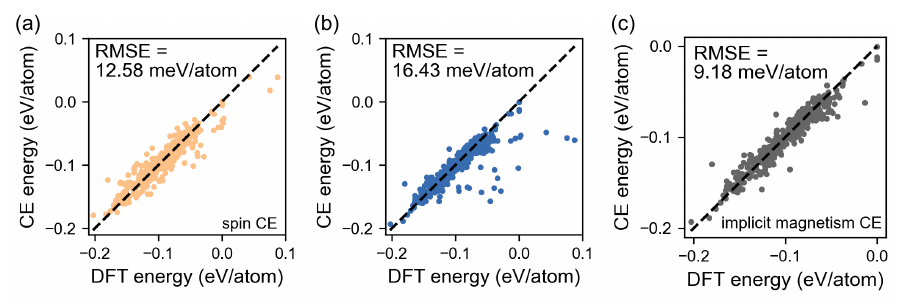}
\caption{ 
Comparison between CE energies and DFT energies for (a) the spin CE fit for all DFT data including magnetic degrees of freedom, (b) the average magnetism CE fit for all DFT data including magnetic degrees of freedom (a structure in different magnetic states has the same features in this fit), and (c) the implicit magnetism CE fit using only the lowest energy structures in the data set.}
\label{fig4}
\end{figure*}

A comparison between the spin CE energies and DFT energies for the best-performing spin CE is shown in Figure~\ref{fig4}(a).
The RMSE is 12.58 meV/atom, similar to previous CE studies~\cite{wrobel2015phase}.
To see how the spin CE differs from other possible CEs, we refitted two comparison models using only the chemical clusters, i.e., 7 dimers, 12 trimers, and 1 quadrumer. 
The first comparison model in Figure~\ref{fig4}(b) was fitted to the whole data set including all spin configurations. 
This fit results in an increased RMSE  to 16.43 meV/atom. 
The second in Figure~\ref{fig4}(c), referred to as the implicit magnetism CE, was fitted to only the magnetic ground state of each structure.
This fit results in a reduced RMSE of 9.18 meV/atom. 

The increased RMSE in Figure~\ref{fig4}(b) arises from underfitting, due to an insufficiently expressive model. 
As expected, the CE underestimates several DFT configuration energies, likely those that are in magnetically excited states.  
In this case, the model is not sufficiently expressive to capture the dependence of the energy on chemical and magnetic configuration. 
A chemical configuration in different magnetic states always contains the same features, making the fit an average magnetism CE model.
In contrast, the implicit magnetism CE results in a reduced RMSE of 9.18 meV/atom (Figure~\ref{fig4}(c)). 
In this comparison model, the model is both fitted to and evaluated on only the magnetic ground state given a chemical configuration. 
While this model achieves the lowest RMSE, as shown below fitting and evaluation in a reduced configuration space (magnetic ground states) leads to poor predictive ability when the model is used outside of that configuration space.  

\begin{figure*}[!hbtp] 
\centering
\includegraphics[width=6in]{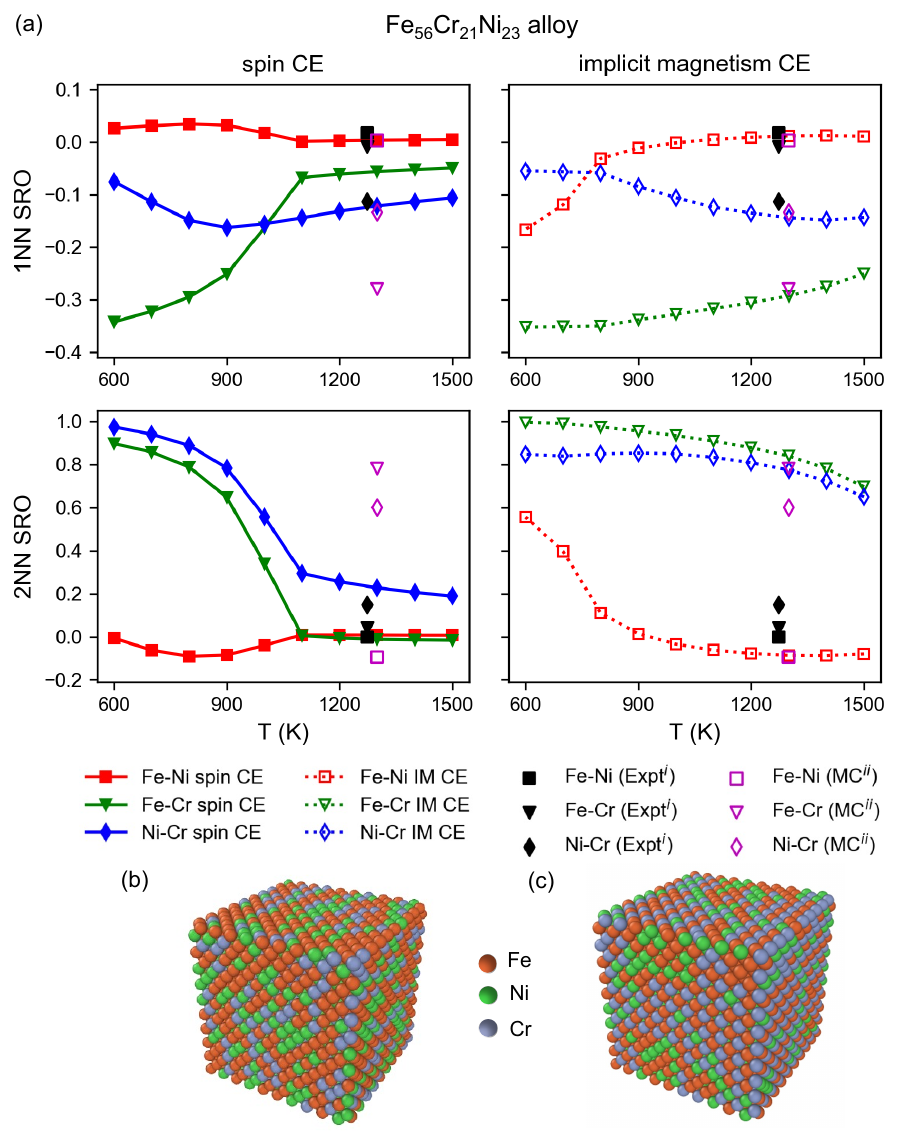}
\caption{
(a) 1 NN and 2 NN SRO parameters of Fe$_{56}$Cr$_{21}$Ni$_{23}$ alloy as functions of temperature. The left panel is from the spin CE (solid lines) and the right panel is from the implicit magnetism (IM) CE (dashed lines). (b) MC snapshot from the spin CE at 1300K. (c) MC snapshot from the implicit magnetism CE at 1300K. Images are generated in OVITO~\cite{stukowski2009visualization}. Fe, Ni, and Cr atoms are marked with red, green, and blue, respectively. 
Note: i. The experimental data is taken from Ref.~\cite{cenedese1984diffuse}.
ii. The MC result is taken from Ref.~\cite{wrobel2015phase}. 
}
\label{fig5} 
\end{figure*}

Figure~\ref{fig5} shows the performance of the spin CE (Figure~\ref{fig4}(a)) and the implicit magnetism CE (Figure~\ref{fig4}(c)) against available experimental benchmarks. 
The SRO parameter of Fe$_{56}$Cr$_{21}$Ni$_{23}$ alloy as a function of temperature is taken from multiple sources. 
Experimental~\cite{cenedese1984diffuse} and theoretical~\cite{wrobel2015phase} results from prior studies are represented by solid and hollow symbols, respectively. 
The spin CE, implicit magnetism CE, and previous CE-MC results 
all agree with experiment for 1 NN SRO parameters for Fe-Ni and Ni-Cr measured at 1300~K. 
For the 1 NN Fe-Cr SRO (triangles), however, both the implicit magnetism CE and previous CE-MC results overestimate the degree of SRO. 
When considering Fe-Cr (triangles) and Ni-Cr (diamonds) 2 NN SRO, the deviations between the implicit magnetism CE and experimental data become more pronounced. 
The current implicit magnetism CE and prior CE-MC results are in agreement with each other (but not experiment), most probably because the prior CE-MC was also fitted to the magnetic ground state.  

By contrast, for all six SRO parameters measured in experiment, the spin CE captures trends qualitatively and often quantitatively.
The main difference between the spin CE and the implicit magnetism CE lies in Fe-Cr and Ni-Cr SRO parameters.  
The implicit magnetism CE is fitted to the magnetic ground state for every chemical configuration, in which Cr atoms tend to have anti-aligned spins with Fe and Ni neighbors. 
However, at high Cr content this preferred ordering may not be possible due to configurational frustration. 
Also, at high temperatures, deviations from the lowest energy magnetic ground state are statistically more probable. 
As a result, the implicit magnetism CE predicts unphysically strong interactions between Cr and other species at high temperatures. 
The spin CE resolves this issue by explicitly incorporating magnetic degrees of freedom, positive Cr-Cr and Fe-Cr 1 NN, as well as Fe-Cr and Ni-Cr 3 NN, spin interactions. 
Particularly at higher temperatures, the capacity to account for thermal disorder of magnetic spins reduces the tendency for Fe-Cr and Ni-Cr ordering. 
Monte Carlo configuration snapshots at 1300 K are shown in Figure~\ref{fig5}(b,c) for the spin CE and implicit magnetism CE, respectively. 
These snapshots show that the implicit magnetism CE predicts a strong Fe-Cr ordering that contradicts experiments at 1300K, while spin CE predicts a more disordered configuration.

\begin{figure*}[!hbtp] 
\centering
\includegraphics[width=4.5in]{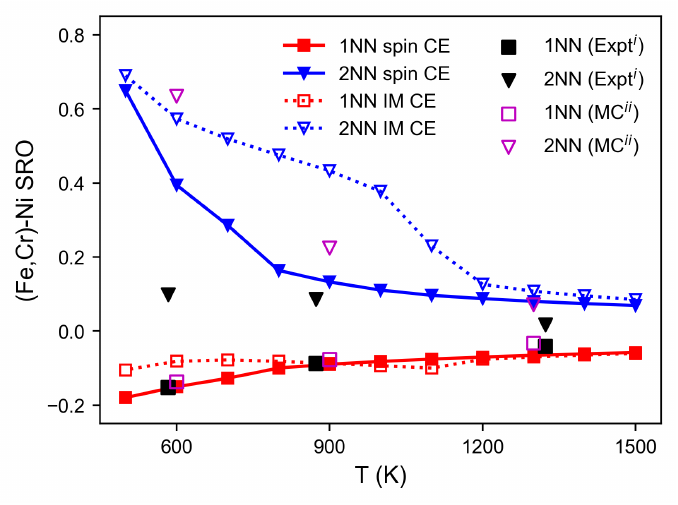}
\caption{
Temperature dependence of 1 NN and 2 NN SRO parameters of Fe$_{42.5}$Cr$_{7.5}$Ni$_{50}$ alloy.
The experimental data are taken from Ref.~\cite{menshikov1997local}. The prior MC results are taken from Ref.~\cite{wrobel2015phase}.
}
\label{fig6} 
\end{figure*} 


Another benchmark, now for  Fe$_{42.5}$Cr$_{7.5}$Ni$_{50}$ alloy~\cite{menshikov1997local}, is shown in Figure~\ref{fig6}.  
Experimental data for the average (Fe, Cr)-Ni SRO parameters were obtained at 1300K, 900K, and 600K for a specimen subjected to radiation at 600 K to accelerate diffusion. 
Again, we compare the spin CE, implicit magnetism CE, and previous MC results to experiments in Figure~\ref{fig6}. 
All models agree with each other and experiment for 1 NN SRO parameters at least qualitatively.
The largest difference between the spin CE and the implicit magnetism CE appears in the 2 NN (Fe,Cr)-Ni SRO parameter: the loss of order occurs at lower temperatures in the spin CE, while the implicit magnetism CE preserves order to higher temperatures. 
At 1300~K, all theoretical predictions are close to experiments for 2 NN SRO. 
When the temperature decreases to 900 K, the implicit magnetism CE deviates for 2 NN SRO, while the spin CE continues to yield reasonable predictions. 
As the temperature further decreases to 600~K, all methods overestimate the 2 NN SRO parameter compared to experiments. 
Although the precise reason for the discrepancy is not known, we expect that irradiated samples at 600 K may show deviations from the equilibrated configurations predicted in MC simulations.
Kinetic factors can act as significant barriers to the development of SRO at intermediate temperatures~\cite{marucco1988effects}, making it challenging to compare fully equilibrated MC results with experimental data.

Additional benchmark comparisons are presented in the SI. SI Figure~\ref{Fe34Fe38} benchmarks the models against experiment for ternary 
Fe$_{34}$Cr$_{20}$Ni$_{46}$ and Fe$_{38}$Cr$_{14}$Ni$_{48}$ alloys.  
Benchmarks of the spin CE for binary alloys are available in SI Figure~\ref{binary}. 
Here for binary FeNi and NiCr alloys at high temperatures, the predicted SRO parameters are again close to experiment. 
For chemical order-disorder transition temperatures, the spin CE yields predictions that are approximately 100 K higher than the experimental values.
In terms of magnetic properties, the Curie temperature prediction from the spin CE is shown in SI Figure~\ref{curie}. 
The model somewhat overpredicts the Curie temperature of Ni compared to experiment.
Although it overestimates the Curie temperatures of FeNi and Ni$_{3}$Fe, it does capture the ferromagnetic nature of these alloys. 

In aggregate, the spin CE provides more realistic predictions than the implicit magnetism CE, particularly in capturing reduced ordering arising from finite-temperature magnetism. 
On the other hand, at low temperatures where ground states are expected to dominate the system, both yield similar results. 
The spin CE offers the advantage of retaining magnetic information from all DFT calculations without the need to focus solely on the magnetic ground states. 

\subsection{SRO Effects in Austenitic Stainless Steels}

\begin{figure*}[!hbtp] 
\centering
\includegraphics[width=4.5in]{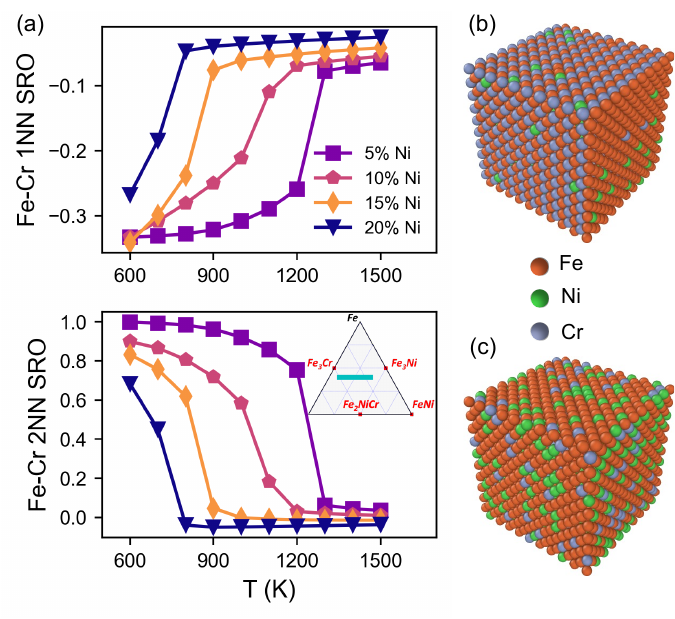}
\caption{ 
(a) The 1 NN and 2 NN Fe-Cr SRO parameters as a function of temperature for Fe-Ni-Cr alloys with different compositions. The inset figure shows a zoomed-in ternary composition space with several stable intermetallic phases highlighted in red squares. The chosen compositions lie on the cyan solid line of the inset figure, where Fe concentration is fixed at 70 \% and Ni concentration changes from 5 \% to 20 \%.
(b) MC snapshot at 900K for the 5 \% Ni case. 
(c) MC snapshot at 900K for the 20 \% Ni case.
Fe, Ni, and Cr atoms are marked with red, green, and blue, respectively.
}
\label{fig7}
\end{figure*} 

\begin{figure*}[!hbtp] 
\centering
\includegraphics[width=6in]{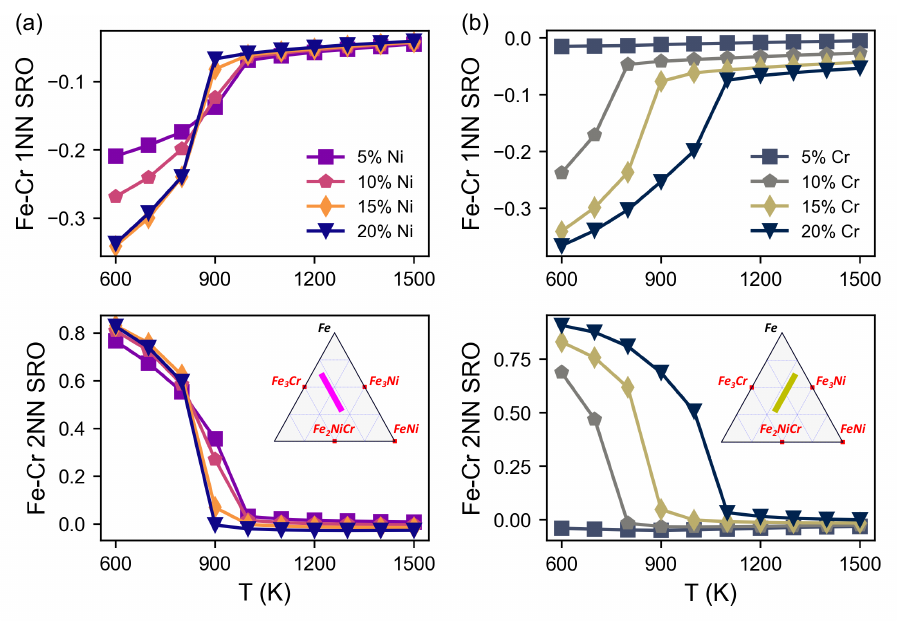}
\caption{
The 1 NN and 2 NN Fe-Cr SRO parameters as a function of temperature for Fe-Ni-Cr alloys with different compositions. (a) The chosen compositions lie on the magenta solid line of the inset figure, where Cr concentration is fixed at 15 \% and Ni concentration changes from 5 \% to 20 \%. (b) The chosen compositions lie on the yellow solid line of the inset figure, where Ni concentration is fixed at 15 \% and Cr concentration changes from 5 \% to 20 \%.
}
\label{fig8} 
\end{figure*}

Having benchmarked the performance of the spin CE across a variety of compositions, we can now use the model to predict the SRO of typical austenitic stainless steels. 
To illustrate the influence of composition on SRO and order-disorder transition temperatures, we calculate the Fe-Cr 1 NN and 2 NN SRO parameters at different temperatures, as shown in Figures~\ref{fig7} and ~\ref{fig8}. 
We fix the concentration of one element and change the concentrations of the other two elements in the composition space of typical austenitic steels (Fe 70\%, Ni 10\%, and Cr 20\%). 
In Figure~\ref{fig7}(a), the Ni concentration is varied from 5\% to 20\%, and the Cr concentration changes accordingly, while the Fe concentration is fixed at 70\%. 
These compositions lie on the cyan line depicted in the inset picture of Figure \ref{fig7}. 
As the Ni/Cr ratio increases, the order-disorder transition temperature decreases significantly from around 1200 K to 700 K. 
With more Ni (and less Cr) present, the degree of Fe-Cr 1 NN and 2 NN SRO accordingly decrease.
Monte Carlo configuration snapshots at 900 K are shown in Figure~\ref{fig7}(b,c) for 5\% and 20\% Ni concentrations, respectively. 
When Ni concentration is 5\%, the alloy shows a strong Fe-Cr ordering at 900K by forming Fe$_3$Cr L1$_2$-like structures.
The Ni atoms appear to be randomly distributed in the lattice.
When the Ni concentration is increased to 20\%, the alloy configuration becomes more disordered and the Fe-Cr ordering is reduced significantly.

We also fix the Cr (Ni) concentration at 15\% and vary the Ni (Cr) concentration from 5\% to 20\% to further investigate the role of Ni and Cr on SRO.  
Figure~\ref{fig8}(a) shows that the transition temperature only decreases slightly when Ni concentration increases at fixed Cr content. 
Here, higher Ni content somewhat promotes ordering tendencies between Fe and Cr below 900K. 
However, when the temperature is higher, SRO remains almost unchanged. 
Varying the degree of Cr present has a more pronounced influence on SRO as shown in Figure~\ref{fig8}(b). 
As the Cr concentration increases, the transition temperature increases significantly, accompanied by a significant increase in SRO at elevated temperatures. 
The addition of Cr in austenitic stainless steels promotes SRO significantly, leading to increased heterogeneity within the alloys. 
For Fe-Ni and Ni-Cr SRO parameters, the effect of the composition is shown in SI Figure~\ref{fix_Fe},~\ref{fix_Cr}, and~\ref{fix_Ni}. 
Similar trends in order-disorder temperature are found for Fe-Ni and Ni-Cr SRO, and Cr again influences SRO most significantly. 
These results show possible routes to control the degree of ordering in austenitic stainless steels, via tailoring of alloy composition and manufacturing/annealing temperatures.

\section{Discussion}


When using theoretical approaches to predict SRO, it is important to describe the material system as realistically as possible. 
For example, the effects of finite temperature magnetism on SRO in complex alloys can be non-negligible.  
We demonstrate that conventional CE models that neglect magnetism, or that only implicitly include magnetism by fitting to magnetic ground state configurations, show discrepancies in SRO of Fe-Ni-Cr alloys compared to experiments.  
The problem lies in the exclusion of magnetic degrees of freedom.
While the implicit approach is effective for capturing configuration energies at low temperatures where the magnetic configurations are in the ground state, it may fail at higher temperatures where the effects of finite-temperature magnetism are non-trivial.
Similarly, the presence of different magnetic states (e.g. above and below the Curie temperature) can affect preferred chemical orderings, so CEs fitted to the magnetic ground state may not be suitable at high temperatures. 
Near magnetic transition temperatures (Curie temperature of Ni is 627 K and of Fe is 1043 K), it is essential to exercise caution when simulating SRO using the CE-MC approach.

There have been previous efforts to address magnetism in the CE method as well.
By employing separate CE models fitted to first-principles energies in different magnetic states, it becomes possible to discern how magnetism affects the pair interactions, consequently altering predictions related to SRO~\cite{woodgate2023interplay}.
This approach focuses on the impact of paramagnetic (PM, corresponding to non-zero but disordered spins) or ferromagnetic magnetic states on \textit{ab initio} energies at 0 K. 
However, a drawback of this method lies in the necessity for separate fits for PM and FM structures, which limits its ability to capture the subtleties of finite-temperature magnetism directly.
This challenge was addressed by \citeauthor{ruban2016atomic}\cite{ruban2016atomic} who considered longitudinal spin fluctuations in \textit{ab-initio} calculations to capture the effects of finite-temperature magnetism.
While this method also provides valuable insights, it introduces temperature- and composition-dependent pair interactions into the CE framework.
This complexity can pose challenges when extending theoretical predictions to diverse alloy systems at varying temperatures.

The spin CE model of this study, on the other hand, provides one coherent fit for both chemical and spin interactions, which does not require separate data sets or fits of different magnetic states. 
The relationship between magnetism and SRO in Fe-Ni-Cr alloys can be studied systematically using the spin CE-MC method. 
For instance, utilizing this model, we validate prior studies on the interplay between chemical order and magnetic order~\cite{dang1996simultaneous,izardar2022impact,ekholm2010influence,walsh2021magnetically,woodgate2023interplay,kollie1973heat,vernyhora2010monte}, and recovered certain experimental findings on SRO~\cite{menshikov1972FeNi,cenedese1984diffuse,menshikov1997local}.
As a result of finite temperature magnetism, we predict that Fe-Cr and Fe-Ni ordering is not as profound as Ni-Cr in the Fe$_{56}$Cr$_{21}$Ni$_{23}$ alloy at high temperatures, consistent with previous experiments~\cite{menshikov1997local}. 
Moreover, we predict here that the presence/absence of SRO in austenitic stainless steels is largely governed by the Cr content of the alloy.  
The dominant influence of Cr content arises from the strong tendency for AFM alignment between Cr and its first nearest neighbors, especially 1 NN Cr-Cr. 
Identification of dominant contributors to SRO (or lack of SRO) can lead to design rules to gain control over SRO, and its consequences for macroscopic deformation modes. 

Further examination of the ways in which magnetism influences SRO and vice versa are presented in SI Figures 15, 16, and 17.
For instance, certain magnetic states can promote SRO dramatically.
Conversely, the chemical order can also affect the magnetic transition behavior: random configurations of Fe-Ni reduce the Curie temperature compared to ordered structures.
The mutual interaction between magnetism and SRO can play an important role in the equilibrium properties of Fe-Ni-Cr austenitic steels.
This might be applied to other transition metal alloys with complex magnetism, which might be important for future studies.

We anticipate that the spin CE could be further improved by increasing the fidelity of first-principles calculations for magnetic interactions, possibly such as non-collinear DFT calculations since spin waves have been reported in related alloys~\cite{hatherly1964spin,shirane1968spin}. 
Additionally, longitudinal spin fluctuations and more refined magnetic models (e.g. Heisenberg model) could be incorporated into the CE method. 


Finally, we emphasize the importance of model validation using available experimental benchmarks in future computational studies of SRO.
In this work, different CE models (spin CE vs implicit magnetism CE) that have similar RMSE or CV scores exhibit substantial variations in predicted thermodynamic quantities obtained by MC simulations.
In particular, we observed that the implicit magnetism CE, fitted and evaluated on a simpler data set, obtains lower CV scores but is less predictive of available experimental benchmarks. 
Consequently, it is imperative to benchmark the models and quantitatively assess the uncertainties for consistency.


\section{Conclusions}

We report a spin CE that combines a conventional chemical CE with a spin Ising model to capture the effects of magnetism on SRO. The main findings are:
\begin{enumerate}
    \item Cr atoms prefer to align antiferromagnetically with neighboring atoms in Fe-Ni-Cr alloys, leading to magnetic frustration when the local Cr concentration is high.
    \item Implicit magnetism CE models that ignore magnetic degrees of freedom tend to overestimate SRO, while the spin CE agrees with experimental results over a broad range of compositions and temperatures.
    \item In austenitic stainless steels, Cr content affects SRO and order-disorder temperatures most significantly compared to Fe and Ni. The addition of Cr promotes SRO and may increase alloy heterogeneity.
    \item Magnetism is a primary factor influencing the degree of SRO in Fe-Ni-Cr alloys. The mutual interaction between magnetism and SRO plays a key role in the equilibrium
    properties of Fe-Ni-Cr austenitic steels, and points to design rules for controlling SRO via alloy chemistry.  
\end{enumerate}

\section{Acknowledgments}
The authors acknowledge support from the US Department of Energy H2@Scale program, through award DE-EE0008832. 
This work was also supported by DOE-NNSA through the Chicago/DOE Alliance Center (DE-NA0003975).
This work used PSC Bridges-2 at the Pittsburgh Supercomputing Center through allocation MAT220011 from the Advanced Cyberinfrastructure Coordination Ecosystem: Services \& Support (ACCESS) program, which is supported by National Science Foundation grants \#2138259, \#2138286, \#2138307, \#2137603, and \#2138296.

\newpage
\bibliography{mybib.bib}

\end{document}


\newpage
\subsection{The formalism of different CE models}

The general form of the conventional CE Hamiltonian is
\begin{equation}
    E(\vec{\sigma}) = \sum_{\omega} m_{\omega} J_{\omega} \langle \Gamma_{\omega'} (\vec{\sigma}) \rangle_{\omega} \hspace{0.5em}.
\end{equation}

For the simplest binary case, each site i can be occupied by two species with occupation variable $\sigma_i = \pm 1$. 
Like an extended Ising Hamiltonian, the cluster function can be expressed as the product of occupation variables:
\begin{equation}
    \Gamma_{\omega} (\vec{\sigma}) = \prod_{i \in \omega} \sigma_i \hspace{0.5em}.
\end{equation}

For ternary systems and beyond (multicomponent system), if site i can host $M_i$ components then the occupation variable $\sigma_i$ can take values from 0 to $M_i-1$, i.e. $\sigma_i = 0, 1, 2$ can represent atom A, B, and C, respectively.
For each site i, a point function $\gamma_{s_i, M_i} (\sigma_i)$ should be used to express the energy dependence of cluster $\omega$ on the occupation variable $\sigma_i$, where $s_i$ can range from 0 to $M_i-1$.
The cluster function is then
\begin{equation}
    \Gamma^{(s)}_{\omega} (\vec{\sigma}) = \prod_{i} \gamma_{s_i, M_i} (\sigma_i) \hspace{0.5em}.
\end{equation}

Now we can see for a cluster $\omega$  containing $|\omega|$ sites in the multicomponent lattice, each site can be assigned with $M_i-1$ different points functions. 
For all other sites outside of the cluster, $\gamma_{0, M_i} (\sigma_i) = 1$ will be used so they won't matter for the cluster of interest.
The sequence of $\{s_1,s_2...s_{|\omega|}\}$ is referred to as decoration $\{s\}$.
For example, a pair cluster in a ternary lattice can have $2^{3-1} = 4$ decorations.

In particular, point function $\gamma_{s_i, M_i} (\sigma_i)$ must satisfy $\gamma_{0,M_i} (\sigma_i) = 1$ and the orthogonality condition:
\begin{equation}
\frac{1}{M_i} \sum^{M_i-1}_{\sigma_i = 0} \gamma_{s_i, M_i} (\sigma_i) \gamma_{t_i, M_i} (\sigma_i)= 
\begin{cases}
    1 & \text{if } s_i = t_i \hspace{0.5em}\\
    0 & \text{otherwise \hspace{0.5em}.} 
\end{cases}
\end{equation}
While there can be various functions satisfying the conditions, the choice of $\gamma_{s_i, M_i} (\sigma_i)$ used in ATAT is
\begin{equation}
\gamma_{s_i, M_i} (\sigma_i) =
\begin{cases}
    1 & \text{if } s_i = 0\\
    -cos(2\pi\left[\frac{s_1}{2}\right] \frac{\sigma_i}{M_i}) & \text{if } s_i > 0 \text{ and odd} \\
    -sin(2\pi\left[\frac{s_1}{2}\right] \frac{\sigma_i}{M_i}) & \text{if } s_i > 0 \text{ and even}
    \end{cases} 
\end{equation}
where [...] denotes the round up operation, e.g. $[\frac{1}{2}]=1$.
For binary case, this is reduced to
\begin{equation}
\gamma_{s_i, 2} (\sigma_i) = 
\begin{cases}
    1 &  \text{if } s_i = 0 \\
    -(-1)^{\sigma_i} & \text{if } s_i = 1
\end{cases}
\end{equation}
where $\sigma_i = 0, 1$. 
One can see clearly that this result is consistent with the extended Ising Hamiltonian we showed above for simple binary systems. In the ternary case, the point functions are orthogonal but not orthonormal.
Extra normalization factors are needed~\cite{wrobel2015}.
The number of possible dimer clusters in a ternary lattice is 3, i.e., $\langle \gamma_1 \gamma_1 \rangle$, $\langle \gamma_1 \gamma_2 \rangle$, and $\langle \gamma_2 \gamma_2 \rangle$.

For the new CE method described in the main text, we exclude the necessity of an orthonormal basis as point functions.
Clusters are decorated explicitly by the chemical species distributed within the lattice.
For a ternary A-B-C system, the number of possible dimer clusters is 6, i.e., AA, AB, AC, BB, BC, and CC.
The trimers and dimers are highly correlated due to the direct decoration of atomic species. 
For example, AA dimer is associated with AAA/AAB/AAC/... trimers linearly.
Since all clusters with different decorations are fitted independently, LASSO tends to select trimers over dimers. This may not be the case when clusters are decorated with non-linear point functions.

\newpage

\begin{figure*}[!hbtp] 
\centering
\includegraphics[width=3in]{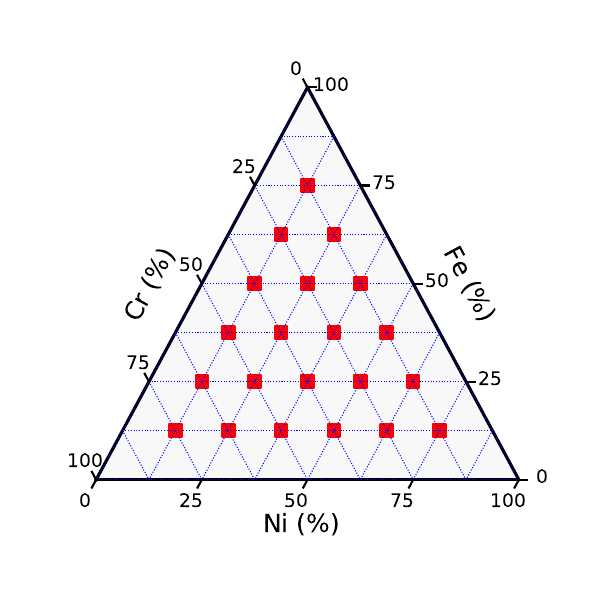}
\caption{
The ternary composition space of Fe-Ni-Cr alloys. The red squares label the SQS compositions sampled in this study. 
}
\label{composition_space} 
\end{figure*} 


\newpage

\begin{figure*}[!hbtp] 
\centering
\includegraphics[width=6in]{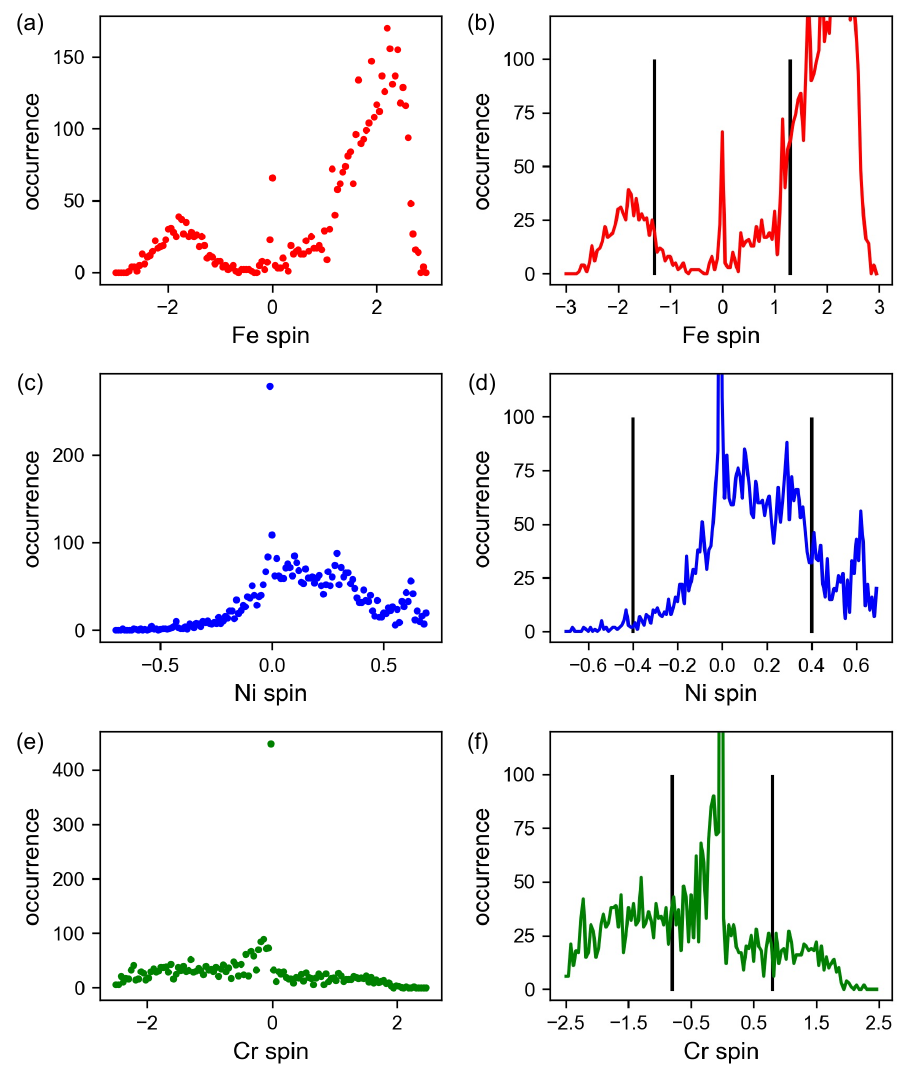}
\caption{
Spin distribution of Fe, Ni, and Cr from the whole data set. (b), (d), and (f) are the zoomed-in versions of (a), (c), and (e), respectively. The black lines indicate the spin thresholds for each species.
}
\label{mag_distrib} 
\end{figure*} 


\newpage

\begin{figure*}[!hbtp] 
\centering
\includegraphics[width=6in]{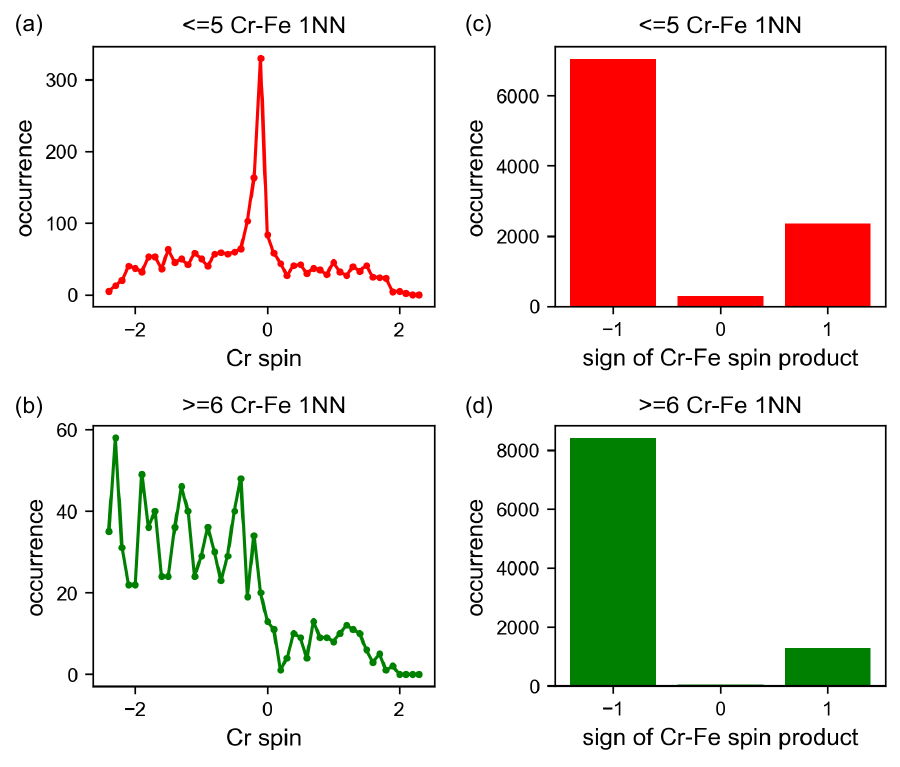}
\caption{ 
(a, b) The distribution of Cr magnetic moment. With more Cr-Fe 1NN pairs, the Cr spins become more negative. (c, d) The distribution of the Fe-Cr spin product. With more Cr-Fe 1NN around Cr atoms, the probability of anti-aligned spins between Cr and Fe is higher.
}
\label{CrFe}
\end{figure*} 


\newpage

\begin{figure*}[!hbtp] 
\centering
\includegraphics[width=6in]{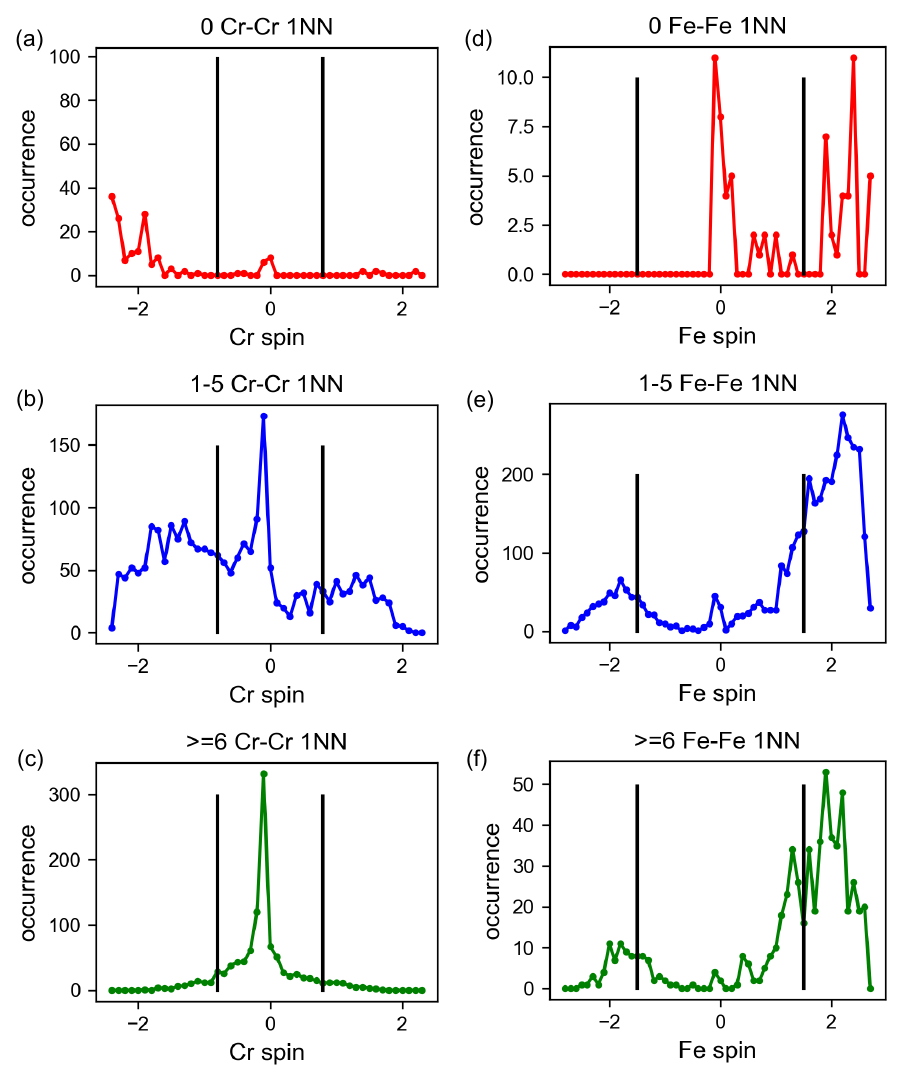}
\caption{
(a-c) The effect of the number of Cr-Cr pairs on Cr spin distribution. (d-f) The effect of the number of Fe-Fe pairs on Fe spin distribution. The black lines indicate the spin thresholds we adopt for Fe and Cr.
}
\label{CrCr} 
\end{figure*} 


\newpage

\subsection{On the spin diversity of the data set}

To construct a reliable spin CE, we need to make sure that the data set includes enough information about both configuration and magnetism. 
The configuration diversity can be achieved by arranging atomic occupations on lattice sites randomly or in order.
However, the spin diversity is difficult to control.
Even when initialized with different magnetic states, many alloy structures eventually relax to similar spin configurations in DFT calculations.
In the dataset we generated, most of the structures have a zero spin product for Ni-Cr pairs.
Will it give rise to the vanished Ni-Cr spin interactions in LASSO?

To address this issue, we duplicated all the structures that contain non-zero spin products of Ni-Cr 1 NN pairs in the data set.
The biased fit on this new data set did give non-zero Ni-Cr spin interactions in contrast to the zero Ni-Cr spin interactions in the normal fit.
However, the biased fit failed to match most of the benchmark tests.

We also weighted the structures that contain non-zero Ni-Cr 1 NN spin products by a factor of two during the LASSO fit.
The Ni-Cr 1 NN spin interactions still vanished in this fit.
The corresponding MC simulation gave similar SRO results compared to the normal fit but failed to predict the magnetic transition of Ni.
We note here that the diversity of spins could be a potential issue for magnetic CE.
However, this might not be the case in our study.

\newpage 

\begin{figure*}[!hbtp] 
\centering
\includegraphics[width=4.5in]{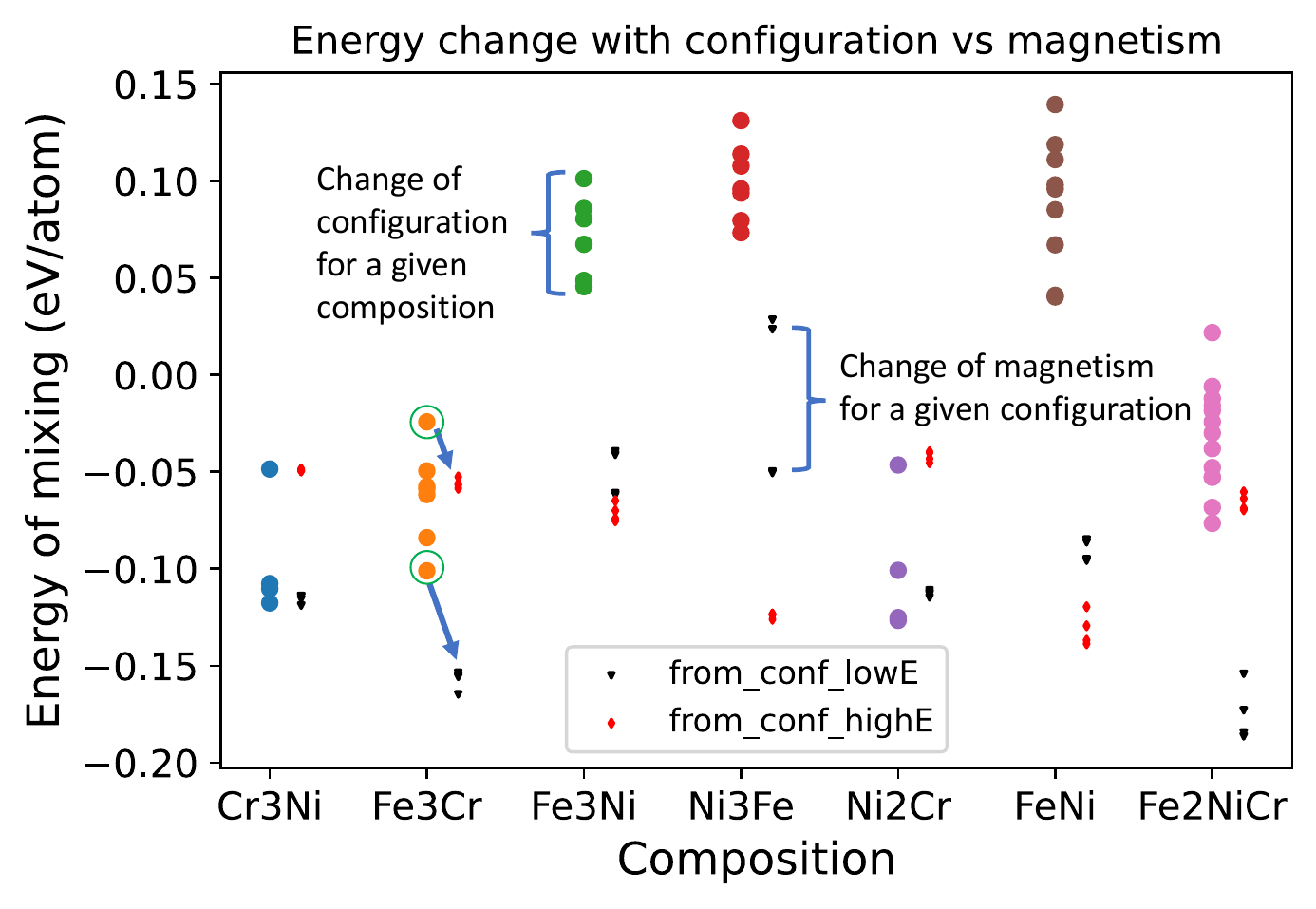}
\caption{
The comparison between magnetic energy and configurational energy for different intermetallic compositions. The large circles indicate the non spin-polarized DFT calculations. The small triangles and diamonds indicate the spin-polarized DFT calculation. The triangles/diamonds are from the lowest/highest energy configurations (in non spin-polarized DFT) with varying magnetic states.
}
\label{mag_vs_conf} 
\end{figure*} 


\newpage

\begin{figure*}[!hbtp] 
\centering
\includegraphics[width=4.5in]{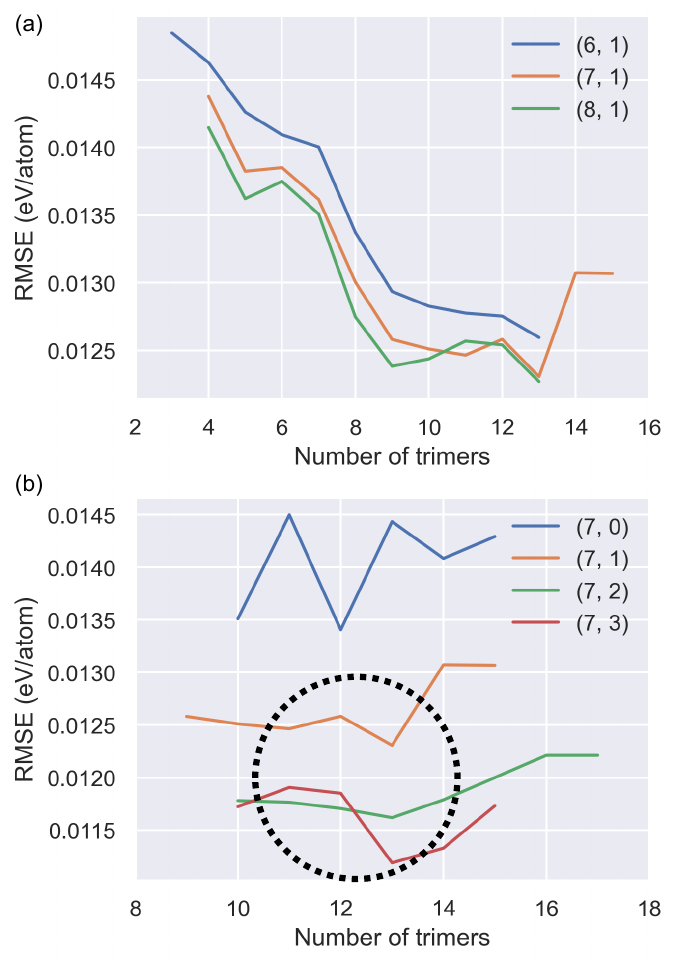}
\caption{
The root mean square error of the LASSO fit as a function of the input number of clusters. The X-axis is the number of trimers. (a) the number of dimers changes from 6 to 8 while the number of quadrumers is fixed as 1. (b) the number of quadrumers changes from 6 to 8 while the number of dimers is fixed as 7.
}
\label{changing_clusters} 
\end{figure*} 


\newpage

\begin{figure*}[!hbtp] 
\centering
\includegraphics[width=6in]{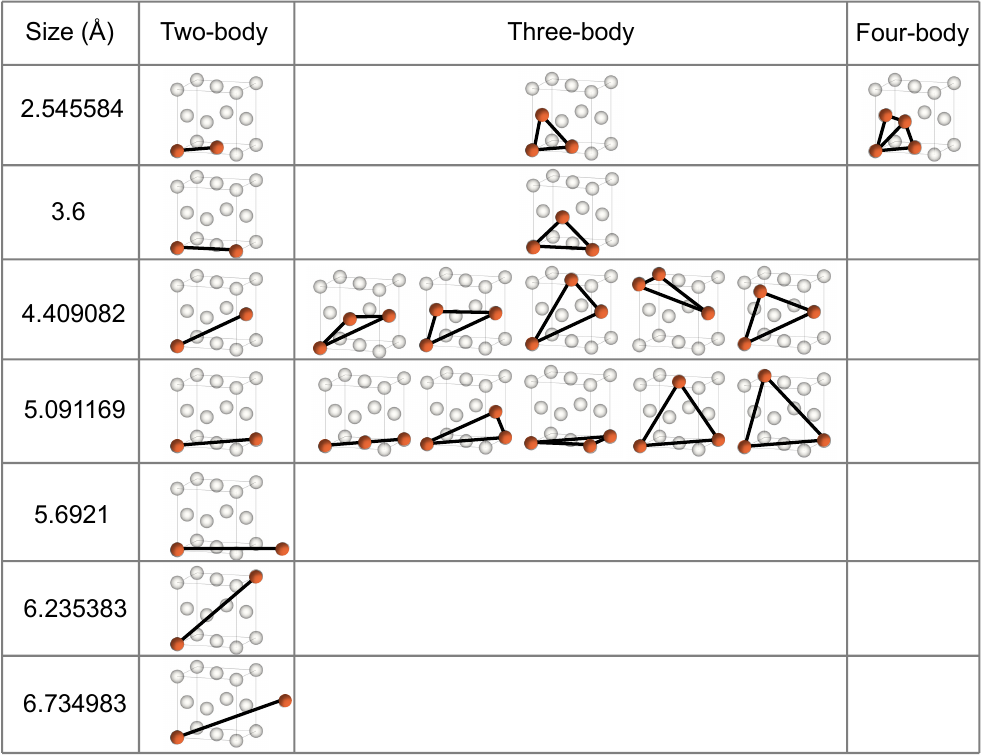}
\caption{
The scheme of all input clusters for the CE construction by LASSO.
For this set of input clusters, the trimer terms represent a complete set encompassing interactions up to the 4th shell. 
Although increasing the number of trimers from 12 to 13 appears to result in a reduced RMSE during the fitting process, the 13$^{th}$ trimer extends into the 5$^{th}$ shell.
To fully encompass all trimers up to the 5$^{th}$ shell, the number of trimers would need to exceed 16, ultimately leading to overfitting as illustrated in SI Figure~\ref{changing_clusters}(a). 
Furthermore, it's worth noting that the model comprising 13 trimers does not align with some benchmarks. 
Consequently, it is important to avoid including a cluster of a specific size while omitting other potential clusters of the same size, as this may not be a sound practice.
A new guideline may be added to the cluster selection rules suggested previously by ~\citeauthor{van2002}~\cite{van2002}, i.e., always include all possible $N$-body clusters within a given size.
}
\label{all_clusters} 
\end{figure*} 



\newpage

\begin{figure*}[!hbtp] 
\centering
\includegraphics[width=3in]{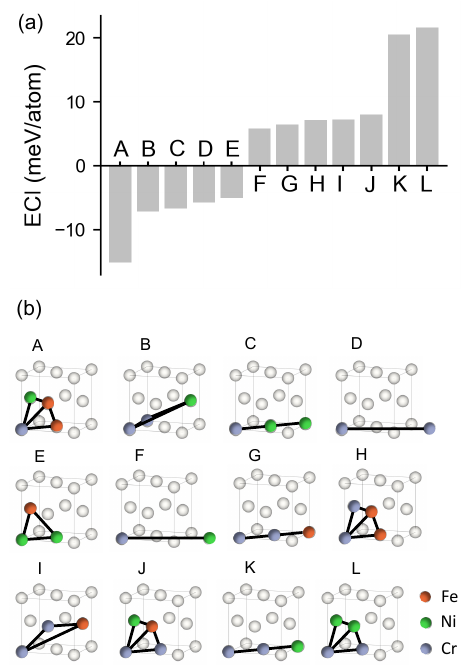}
\caption{
The important chemistry ECIs from implicit magnetism CE (conventional CE) with magnitudes larger than 5 meV. The largest two (positive) ECIs also appear in the selected chemical clusters of the spin CE, but the magnitudes here are two times larger than those in the spin CE. These large magnitudes of the interaction between Cr and other atoms might lead to overestimating Cr-related SRO parameters based on the implicit magnetism CE.
}
\label{eci_nonmag} 
\end{figure*} 


\newpage


\newpage

\begin{figure*}[!hbtp] 
\centering
\includegraphics[width=4.5in]{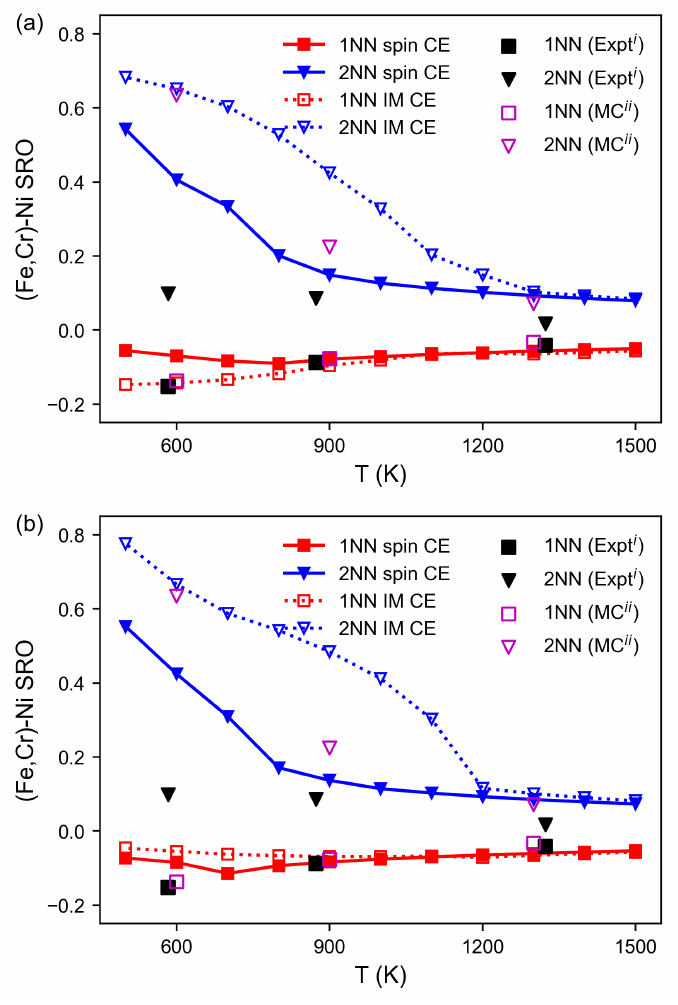}
\caption{
The averaged SRO of (a) Fe$_{34}$Cr$_{20}$Ni$_{46}$ alloy and (b) Fe$_{38}$Cr$_{14}$Ni$_{48}$ alloy. 
Note: i. The experimental data is taken from Ref.~\cite{menshikov1997}.
ii. The MC result is taken from Ref.~\cite{wrobel2015}.
}
\label{Fe34Fe38} 
\end{figure*} 


\newpage

\begin{figure*}[!hbtp] 
\centering
\includegraphics[width=6in]{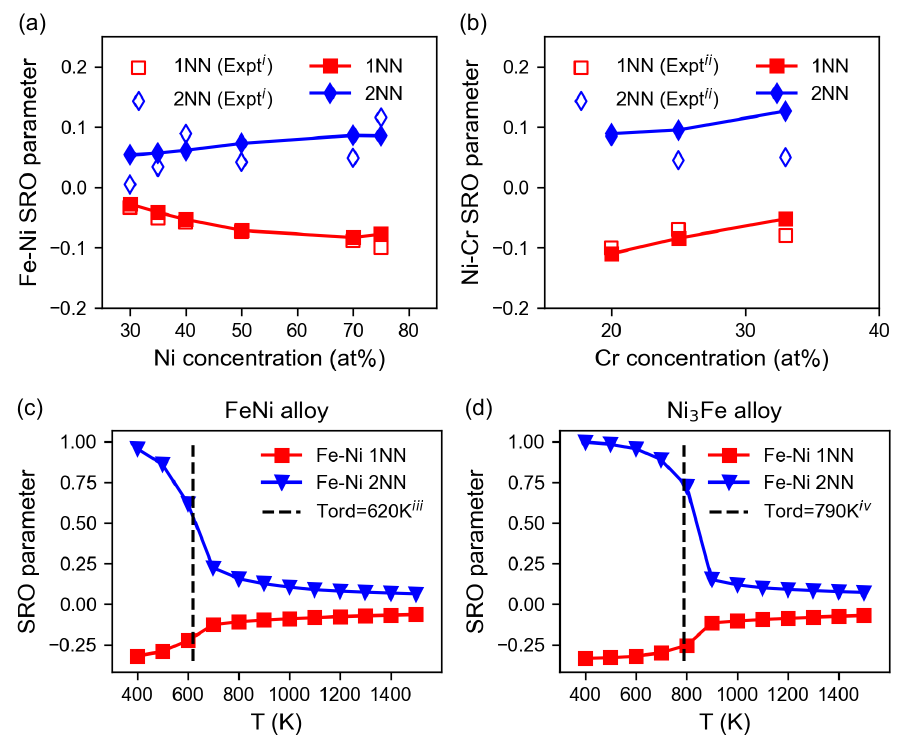}
\caption{
(a) The 1NN and 2NN Fe-Ni SRO for binary Fe-Ni alloys at 1297K. (b) The 1NN and 2NN Ni-Cr SRO for binary Ni-Cr alloys at around 1000K. (c) The order-disorder transition for FeNi. (d) The order-disorder transition for Ni$_3$Fe.
Note: i. The experimental data is taken from Ref.~\cite{goman1971,menshikov1972}.
ii. The experimental data is taken from Ref.~\cite{caudron1992,schonfeld1988}.
iii. The experimental data is taken from Ref.~\cite{mandal2022}.
iv. The experimental data is taken from Ref.~\cite{massalski1990}.
}
\label{binary} 
\end{figure*} 


\newpage

\begin{figure*}[!hbtp] 
\centering
\includegraphics[width=3in]{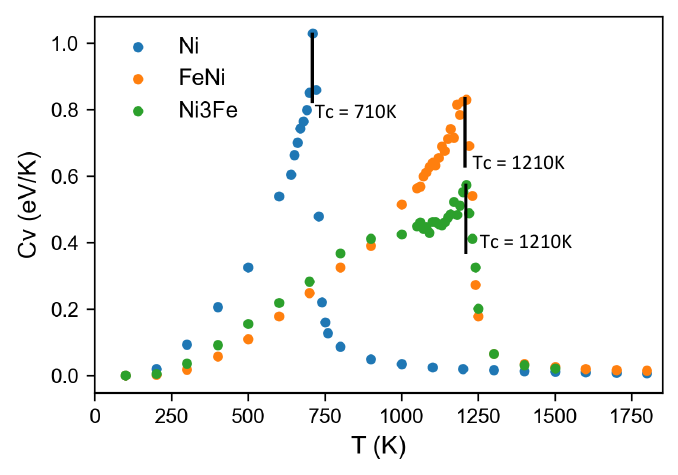}
\caption{
The Curie temperatures of Ni, Ni$_3$Fe, and FeNi are 710K, 1210K, and 1210K, respectively. The experimental Curie temperatures for Ni, FeNi, and Ni$_3$Fe are 627K, 823K, and 853K, respectively~\cite{chicinas2002,wasilewski1988}. 
Note that the Ising model tends to predict a higher magnetic transition temperature compared to the Heisenberg model using the same set of interactions~\cite{domb1962}.
This overestimate arises because the energy barrier to flip one spin to the opposite direction is lower in the Heisenberg model which allows successive small rotations of spins.
}
\label{curie} 
\end{figure*} 


\newpage

\begin{figure*}[!hbtp] 
\centering
\includegraphics[width=6in]{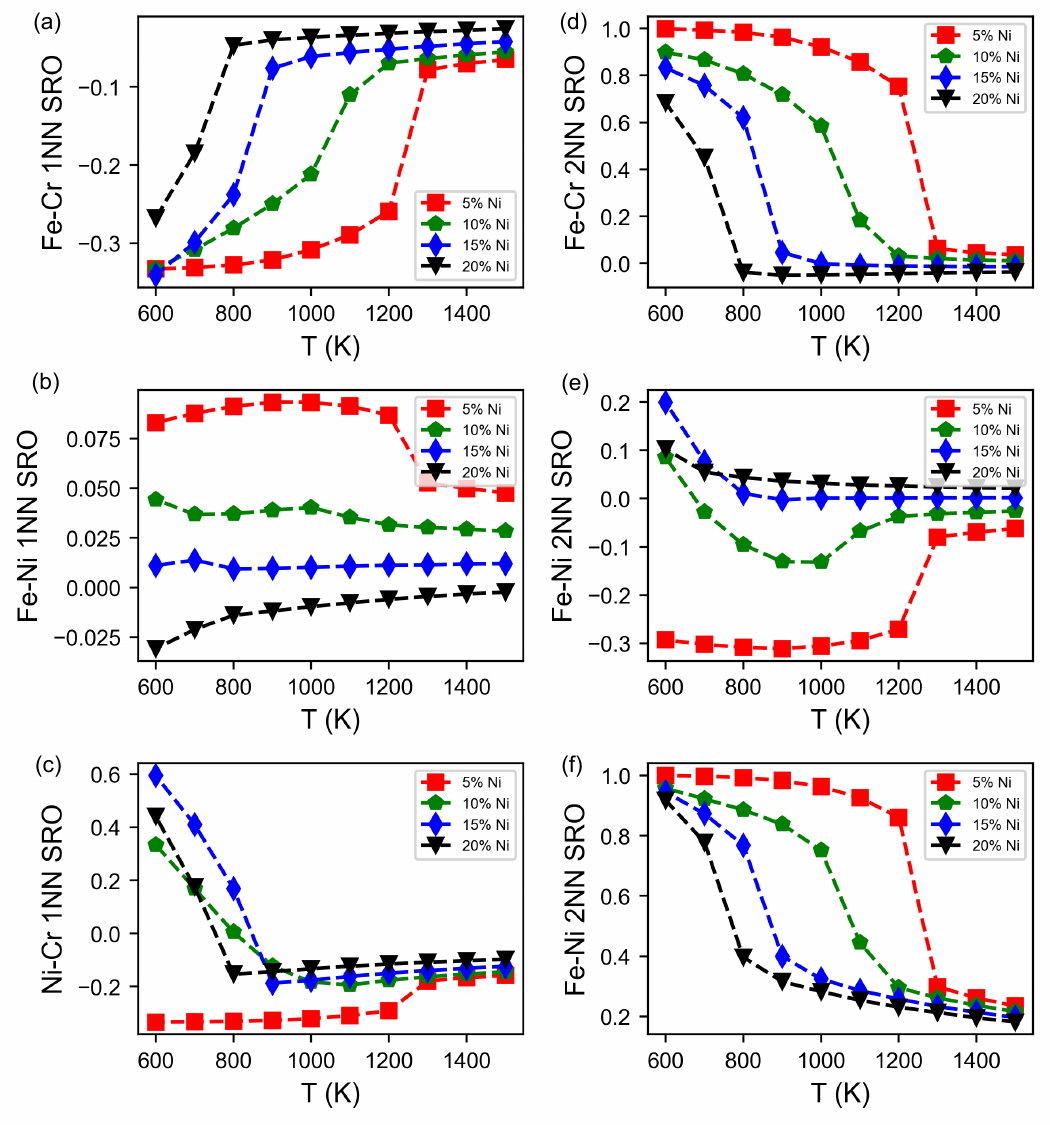}
\caption{
(a-c) 1 NN and (d-e) 2 NN Fe-Cr, Fe-Ni, and Ni-Cr SRO parameters as functions of temperature for Fe-Ni-Cr alloys with different compositions. Here Fe concentration is fixed at 70 \%. Ni concentration changes from 5 \% to 20 \% while Cr content changes accordingly.
}
\label{fix_Fe} 
\end{figure*} 

\newpage

\begin{figure*}[!hbtp] 
\centering
\includegraphics[width=6in]{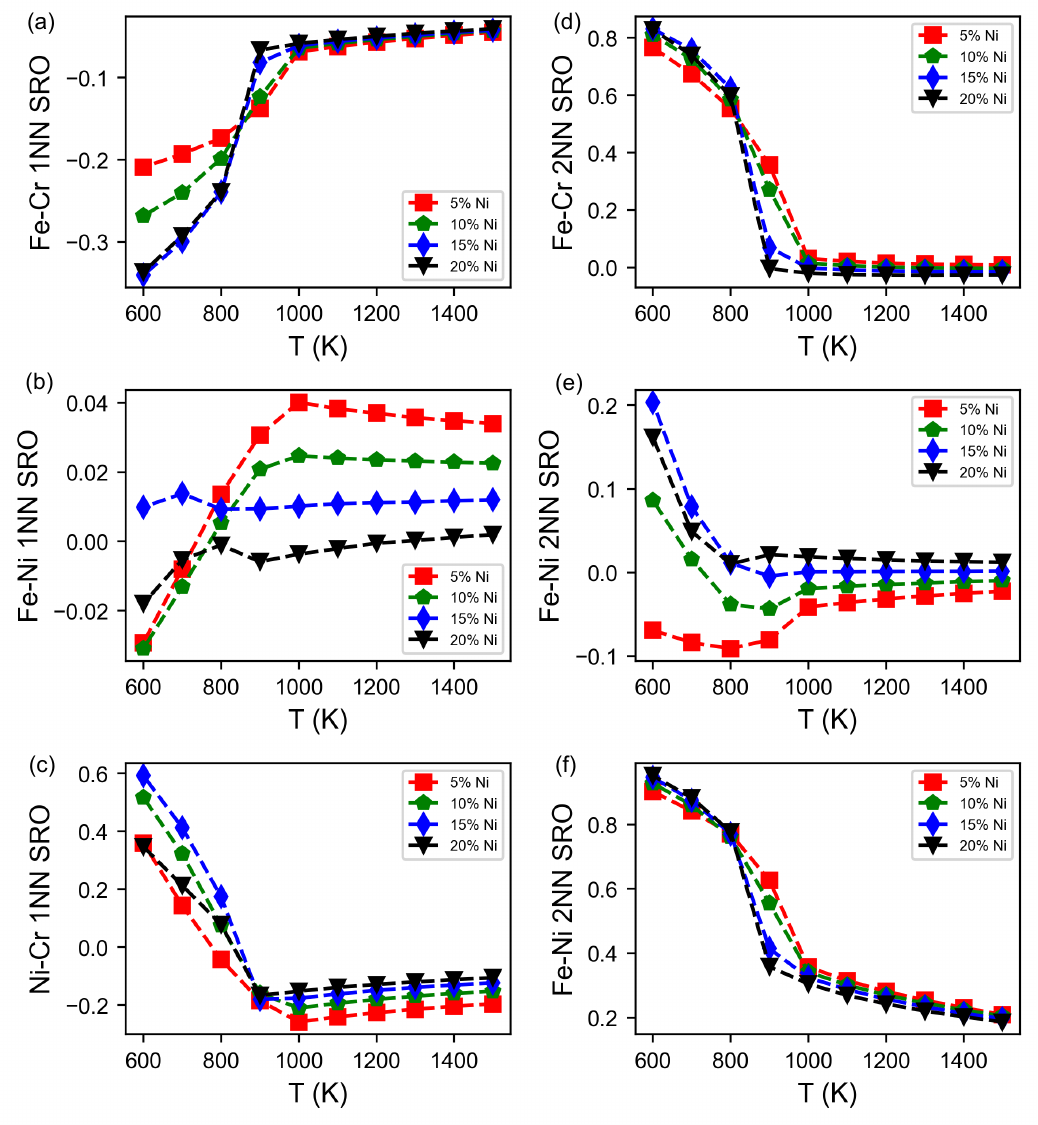}
\caption{ 
(a-c) 1 NN and (d-e) 2 NN Fe-Cr, Fe-Ni, and Ni-Cr SRO parameters as functions of temperature for Fe-Ni-Cr alloys with different compositions. Here Cr concentration is fixed at 15 \%. Ni concentration changes from 5 \% to 20 \% while Fe content changes accordingly.
}
\label{fix_Cr}
\end{figure*} 

\newpage

\begin{figure*}[!hbtp] 
\centering
\includegraphics[width=6in]{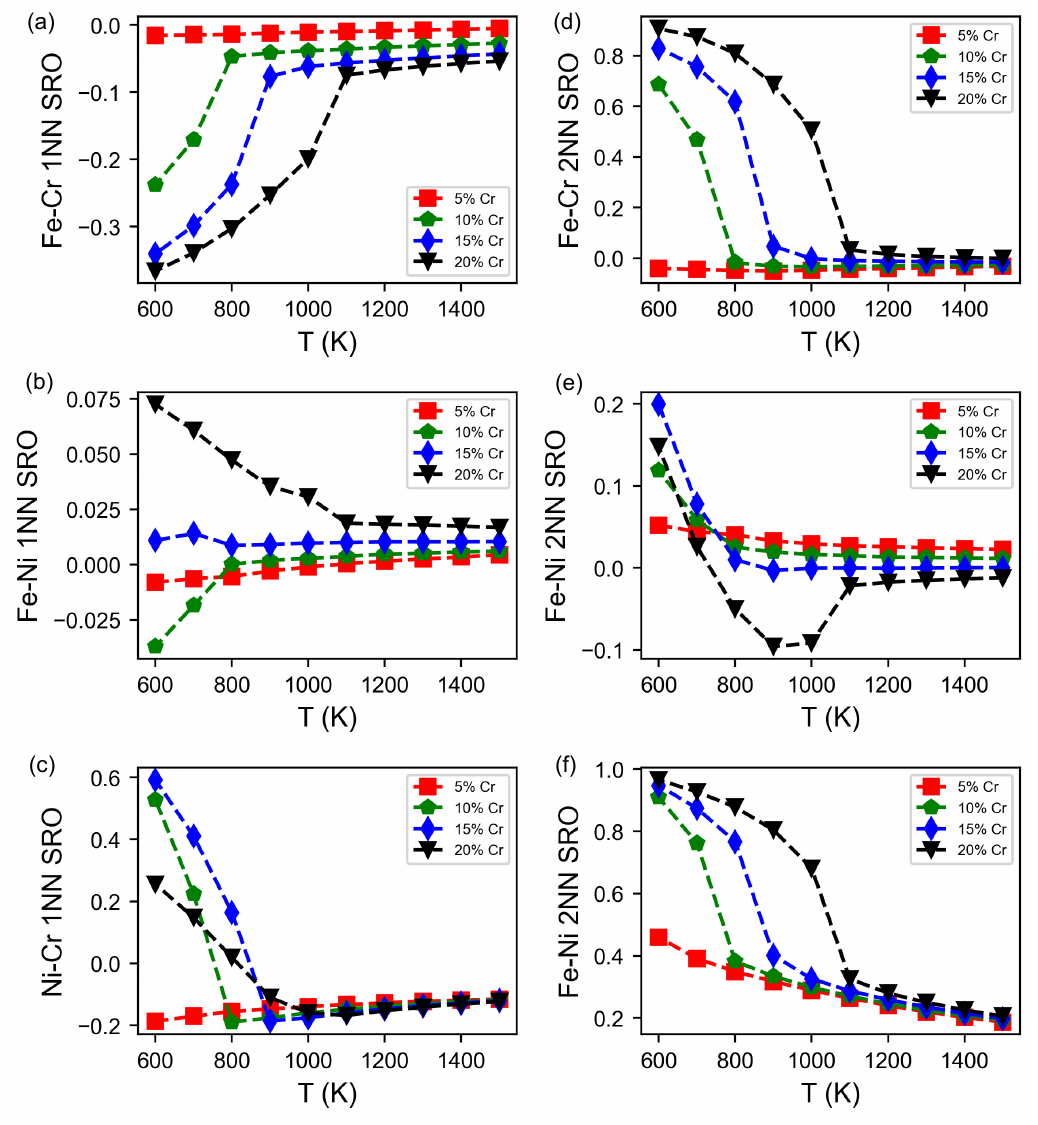}
\caption{
(a-c) 1 NN and (d-e) 2 NN Fe-Cr, Fe-Ni, and Ni-Cr SRO parameters as functions of temperature for Fe-Ni-Cr alloys with different compositions. Here Ni concentration is fixed at 15 \%. Cr concentration changes from 5 \% to 20 \% while Fe content changes accordingly.
}
\label{fix_Ni} 
\end{figure*} 

\newpage

\subsection{The interplay between SRO and magnetism}

\begin{figure*}[!hbtp] 
\centering
\includegraphics[width=6in]{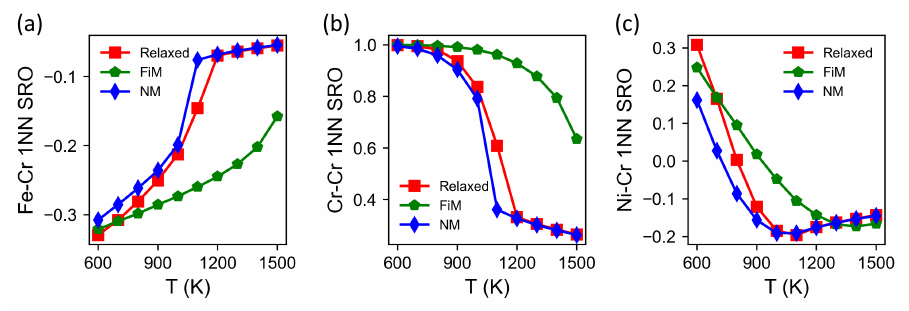}
\caption{
(a) Fe-Cr, (b) Cr-Cr, and (c) Ni-Cr 1 NN SRO parameters as functions of temperature for Fe$_{70}$Cr$_{20}$Ni$_{10}$ alloy in varying magnetic states. 
}
\label{varying_mag} 
\end{figure*}

In SI Figure~\ref{varying_mag} we present the effect of magnetism on SRO in Fe-Cr-Ni austenitic stainless steels by employing the spin CE.
To model typical austenitic steels, we set the alloy composition to Fe$_{70}$Cr$_{20}$Ni$_{10}$.
We compared three cases within CE-MC: a fully equilibrated magnetic state, a non-magnetic (NM) state where all spins were constrained to be zero, and an enforced ferrimagnetic (FiM)-like state with spin-up for Fe and Ni and spin-down for Cr. 
For Fe-Cr 1 NN SRO (SI Figure \ref{varying_mag}(a)), the enforced FiM state results in a higher (more negative) ordering tendency at elevated temperatures compared to the other two cases. 
A similar trend can be seen for Cr-Cr SRO (SI Figure \ref{varying_mag}(b)), where the increased positive SRO parameter indicates that Cr-Cr 1 NN pairs are less favored.
However, for Ni-Cr nearest neighbors (SI Figure \ref{varying_mag}(c)), the influence of the magnetic state is small with the equilibrated, FiM, and NM configurations showing similar degrees of SRO.

The trends observed in SI Figure 7 can be explained in terms of the expected magnetic tendencies of each element. 
For example, the large positive value of the Cr-Cr 1 NN spin ECI promotes a strong AFM state for Cr atoms. 
When the FiM state is enforced, Fe-Cr pairs are favorable as Fe and Cr always have anti-parallel spins, which is exemplified by the negative SRO parameter in SI Figure \ref{varying_mag}(a). 
Meanwhile, Cr-Cr pairs are highly unfavorable as a result of geometric frustration, resulting in large positive SRO parameters across the full temperature regime in SI Figure \ref{varying_mag}b. 
On the other hand, the Ni-Cr spin ECI are comparatively small in magnitude, so the magnetic effect on Ni-Cr SRO in SI Figure \ref{varying_mag}(c) when FiM ordering is imposed is not as dramatic.  
These results highlight artifacts of CE-MC approaches that only fit the magnetic ground state of Fe-Ni-Cr alloys, namely overestimation of both Fe-Cr ordering and Cr-Cr anti-ordering tendencies, arising when the magnetic ground states are imposed. 

\begin{figure*}[!hbtp] 
\centering
\includegraphics[width=3in]{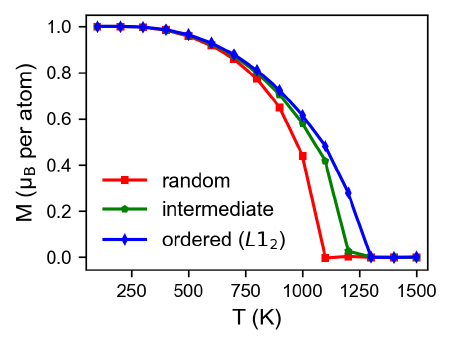}
\caption{
The magnetic transition of Ni$_{3}$Fe alloy with different ordering tendencies from completely random to ordered $L1_2$ structures. }
\label{varying_sro} 
\end{figure*} 


Beyond using the spin CE to investigate how magnetism affects SRO, inversely we can use the model to study the effect of chemical disorder on the magnetic transition temperatures. 
The average magnetic moment of Ni$_{3}$Fe alloy as a function of temperature is shown in SI Figure~\ref{varying_sro} for alloys with different degrees of ordering.
As the alloy structures transition from $L1_{2}$ ordered to fully random configurations, the Curie temperature decreases by around 200 K.
This phenomenon, where chemical order leads to an increase in the magnetic transition temperature, has been verified both experimentally~\cite{kollie1973} and theoretically~\cite{vernyhora2010} in prior studies.
The variation in the Curie temperature arises because AFM Fe-Fe 1 NN magnetic interactions are minimized as the Ni$_{3}$Fe alloy transitions into the $L1_{2}$ ordered state, thus promoting the increase of Curie temperature. 
Another example in which chemical order affects the magnetic transition was put forward by ~\citeauthor{izardar2022}.
They showed that Fe-Fe magnetic exchange interaction in Fe-Ni alloys is influenced by chemical order or local coordination environment~\cite{izardar2022}, giving rise to large variations of magnetic interactions and a reduction of Curie temperature.

\subsection{The phase stability origin of SRO and the role of magnetism}






The thermodynamic driving force associated with the formation of SRO has been correlated with the stability of the random solid solution phase relative to competing ordered phases. 
\begin{figure*}[!hbtp] 
\centering
\includegraphics[width=6in]{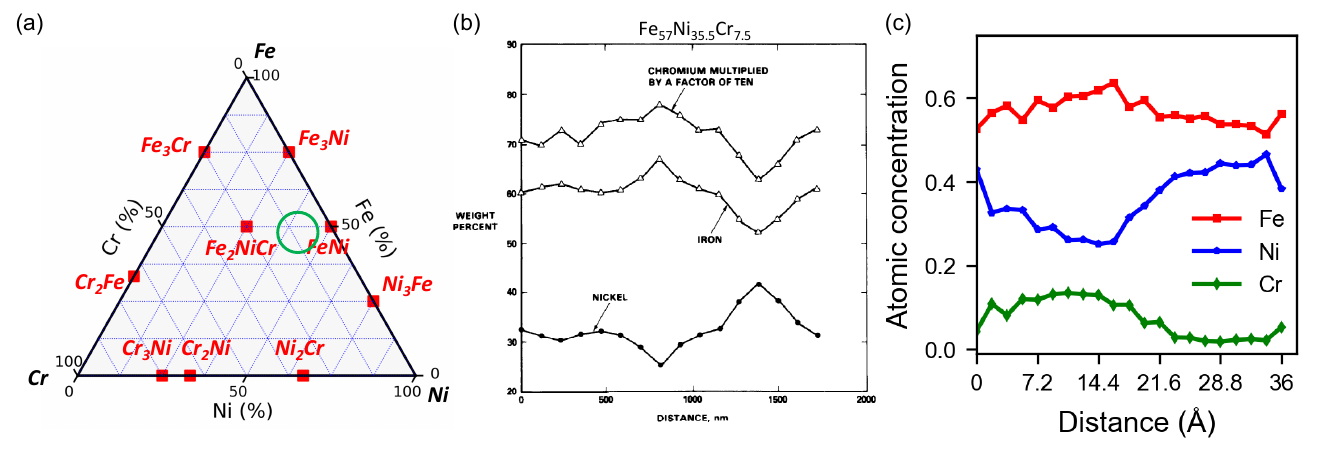}
\caption{ 
(a) the phase stability of FCC Fe-Ni-Cr alloys with several important intermetallic phases labeled in red squares. The circular region contains the composition of Fe$_{57}$Ni$_{35.5}$Cr$_{7.5}$. (b) Experimental concentration profile of the Fe$_{57}$Ni$_{35.5}$Cr$_{7.5}$ alloy at 873K, adapted from Ref.~\cite{garner1993} with permission from Elsevier. (c) Simulated concentration profile in the [100] direction of the Fe$_{57}$Ni$_{35.5}$Cr$_{7.5}$ alloy at 600K using the spin CE. 
}
\label{conc_profile}
\end{figure*} 
To illustrate the relationship between phase stability and SRO, SI Figure~\ref{conc_profile}(a) shows the ternary Fe-Ni-Cr composition space, with stable ordered phases (obtained from DFT simulations) indicated. 
The Fe$_{57}$Ni$_{35.5}$Cr$_{7.5}$ alloy are within the circular region where the metastable austenitic ternary alloys may undergo phase decomposition into various intermetallic phases, such as Fe$_3$Ni, FeNi, and Fe$_2$NiCr.
SI Figure~\ref{conc_profile}(b) shows previously reported concentration profiles of the alloy annealed at 873K~\cite{garner1993} 
The concentration profiles are non-uniform, showing heterogenous variations. 
The profiles for Ni and Fe indicate regions that are relatively Ni-depleted and Fe-rich (around 800 nm) and regions that are relatively Ni-rich and Fe-poor (around 1500 nm). 
These regions suggest that there may be Fe$_{3}$Ni and FeNi -like regions present in the alloy, which is consistent with the phase stability analysis. 

In analogy to spinodal decomposition, SRO can appear in the form of local concentration fluctuations.
At high temperatures, the length scale of these fluctuations is relatively small and challenging to detect experimentally.
However, as the temperature decreases, these features evolve into larger length scales at thermal equilibrium, as the SRO that is present transitions into long-range order (LRO) with more obvious local concentration fluctuations. 
SI Figure~\ref{conc_profile}(c) shows an example of a concentration profile obtained from a MC snapshot generated by the spin CE for  Fe$_{57}$Ni$_{35.5}$Cr$_{7.5}$ alloy.
The simulation cell size is around 4 nm and concentration fluctuations exist in the [100] direction at around 600 K. 
Around 30 \AA, the Fe and Cr concentration decreases simultaneously as the local Ni concentration increases, consistent with the experimental result.
Although the simulation length scale is limited, it does suggest that micro-domain SRO may form to promote concentration fluctuations~\cite{owen2016}, and the driving force for this compositional self-organization is phase stability. 

In SI Figure~\ref{mag_vs_conf}, we have demonstrated that the total energies decrease significantly in spin-polarized calculations compared to non spin-polarized calculations for compositions like FeNi and Ni$_3$Fe.
On the other hand, this change in energy is negligible for compositions like Ni$_2$Cr.
Since magnetic interactions contribute to stabilizing certain alloy structures, the prediction of phase stability might be different if magnetism is neglected.
Although prior studies argued that magnetism is not responsible for the ground-state ordering~\cite{ghosh2022}, we emphasize here that magnetism may affect the ground-state phase stability, thus altering the prediction of SRO in Fe-Ni-Cr alloys.


\newpage
\bibliography{sibib.bib}